\documentclass[sigconf, review=false, anonymous=false, authorversion]{acmart}

\let\oldhref\href
\renewcommand{\href}[2]{\oldhref{#1}{\hbox{#2}}}

\usepackage[utf8]{inputenc}

\usepackage[T1]{fontenc}
\usepackage{amsmath}

\usepackage{bm}

\usepackage{multirow}
\usepackage{subfigure}
\usepackage{listings}

\usepackage[binary, textstyle, amssymb, derived]{SIunits}

\copyrightyear{2022}
\acmYear{2022}
\setcopyright{acmlicensed}
\acmConference[ICPE '22] {Proceedings of the 2022 ACM/SPEC International Conference on Performance Engineering}{April 9--13, 2022}{Bejing, China.}
\acmBooktitle{Proceedings of the 2022 ACM/SPEC International Conference on Performance Engineering (ICPE '22), April 9--13, 2022, Bejing, China}
\acmPrice{15.00}
\acmISBN{978-1-4503-9143-6/22/04} 
\acmDOI{10.1145/3489525.3511689}

\newcommand{\num}[1]{$#1$}
\newcommand{\mycite}[1]{\textit{``#1''}}

\newcommand{\figref}[1]{\figurename{~\ref{fig:#1}}}
\newcommand{\tabref}[1]{Table~\ref{tab:#1}}
\newcommand{\secref}[1]{Section~\ref{sec:#1}}
\newcommand{\lstref}[1]{Algorithm~\ref{alg:#1}}

\title{Memory Performance of AMD EPYC Rome and Intel Cascade Lake SP Server Processors}

\author{Markus Velten}
\email{{givenname.lastname}@tu-dresden.de}
\author{Robert Sch\"{o}ne}
\author{Thomas Ilsche}
\author{Daniel Hackenberg}
\affiliation{
    \institution{Technische Universit\"{a}t Dresden, Center for Information Services and High Performance Computing (ZIH)}
    \postcode{01062}
    \city{Dresden}
    \country{Germany}}
    
\keywords{AMD Zen 2; AMD EPYC Rome; Intel Xeon Cascade Lake; Intel Xeon Skylake; cache coherence; memory hierarchy}

\begin{document}

\ccsdesc[500]{Computer systems organization~Multicore architectures}
\ccsdesc[300]{Computer systems organization~Single instruction, multiple data}
\ccsdesc[300]{Computer systems organization~Interconnection architectures}

\begin{abstract}
Modern processors, in particular within the server segment, integrate more cores with each generation.
This increases their complexity in general, and that of the memory hierarchy in particular.
Software executed on such processors can suffer from performance degradation when data is distributed disadvantageously over the available resources.
To optimize data placement and access patterns, an in-depth analysis of the processor design and its implications for performance is necessary.
This paper describes and experimentally evaluates the memory hierarchy of AMD EPYC Rome and Intel Xeon Cascade Lake SP server processors in detail.
Their distinct microarchitectures cause different performance patterns for memory latencies, in particular for remote cache accesses. 
Our findings illustrate the complex NUMA properties and how data placement and cache coherence states impact access latencies to local and remote locations.
This paper also compares theoretical and effective bandwidths for accessing data at the different memory levels and main memory bandwidth saturation at reduced core counts. 
The presented insight is a foundation for modeling performance of the given microarchitectures, which enables practical performance engineering of complex applications.
Moreover, security research on side-channel attacks can also leverage the presented findings.
\end{abstract}

\maketitle

\section{Introduction}
\label{sec:intro}
\begin{figure}[t]
 \centering
 \includegraphics[width=\columnwidth]{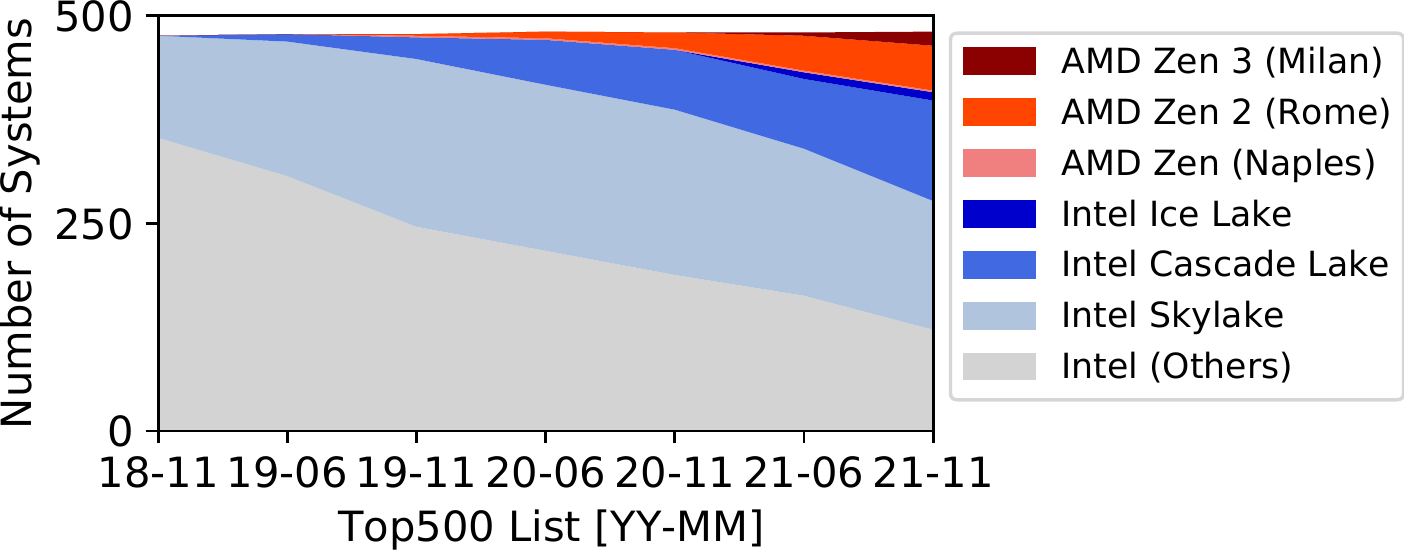}
 \caption{Stacked line chart for Intel and AMD microarchitectures found in Top500 systems~\cite{Top500}}
 \label{fig:top500}
 \Description{Stacked area chart that visualizes the number of AMD Zen, AMD Zen 2, AMD Zen 3, Intel Ice Lake, Intel Cascade Lake, Intel Skylake and other Intel processors within the Top500 list between November 2018 and June 2021.}
\end{figure}

x86 processors dominate the HPC market, with \num{483} systems in the November 2021 Top500 list~\cite{Top500}.
While most of these systems (\num{408}) are Intel-based, AMD continuously increases its market share in recent years (\figref{top500}).
Processors of these two vendors mostly share the same instruction set architecture (ISA), but feature different internal architectures.
The differences have strong implications on the performance of executed codes.
Some criteria such as frequencies, SIMD widths, cache sizes, and number of cores are featured in high level specification descriptions.
Other features are much less prominently documented.
This includes the internal network, which connects cores, DRAM, and I/O, but also the implementation of the cache coherence protocol, which directly influences the memory-bandwidth and latencies under different conditions.
This paper reveals data on these aspects for recent processors: AMD EPYC Rome (\emph{Rome}, implementing the Zen~2 microarchitecture) and Intel Cascade Lake SP (\emph{CLX}, implementing the Cascade Lake microarchitecture).
This information is crucial for various fields, e.g., security~\cite{Yarom2014} and optimization of parallel software~\cite{Sabela2013}.

The remainder of this paper is structured as follows: we introduce related research in \secref{rel} and discuss the Rome and CLX architectures in \secref{rome-arch} and \secref{CLX-arch}, respectively.
\secref{test-system} describes the measurement setup and \secref{local} presents evaluations of local memory accesses.
This is followed by a performance analysis for accesses to remote cache and memory locations within a socket in \secref{intra-socket} and to a remote socket in \secref{inter-socket}.
We summarize our findings and conclude this paper in \secref{conclusion}.

\newpage
\section{Related Work}
\label{sec:rel}

Manuals from the processor vendors provide high level overviews, e.g., AMD's Processor Programming Reference~\cite{AMDRomeRR2020} and the Intel Specification Updates~\cite{IntelSkylakeSpec,IntelCLXSpecUpdate}.
This information is complemented by the respective software optimization guidelines~\cite{AMDOptimization2020,IntelOptimization2020} and additional articles, e.g., on Intel Skylake SP (SKX)~\cite{MulnixSLX2017}.
Further details are presented by processor designers in peer-reviewed articles, which also describe physical implementations and control loops.
Suggs et al. present the Zen~2 architecture in~\cite{Suggs2019, Suggs2020}.
Naffziger et al. describe how multiple dies form a chiplet in~\cite{Naffziger2020Paper, Naffziger2021}.
Tam et al.~\cite{Tam2018} and Arafa et al.~\cite{ArafaCLX2019} detail the SKX and CLX architecture, respectively.

Researchers investigate finer details and independently validate specific details of x86 processors.
The memory subsystem is the main focus of several publications.
Molka et al. cover cache coherence and memory performances of older architectures in~\cite{MolkaNehalem,MolkaSandybridge,MolkaHaswell}.
We continue and extend their work in this paper and compare the more recent architectures Rome and CLX.
Alappat et al. investigated the CLX architecture in~\cite{Alappat2020}.
We extend the research with a more in-depth analysis of the memory latencies, validate the bandwidth results with different benchmarks and compare latency and bandwidth results with Rome.
While cache and memory performance is influenced by power saving mechanisms~\cite{Hackenberg2015_Haswell,romeee,Schoene2019}, the focus of this paper is on the performance at constant processor frequencies.

\section{The AMD EPYC Rome Architecture}
\label{sec:rome-arch}
\subsection{General Concept}
\begin{figure}[p]
    \centering
    \includegraphics[width=\columnwidth]{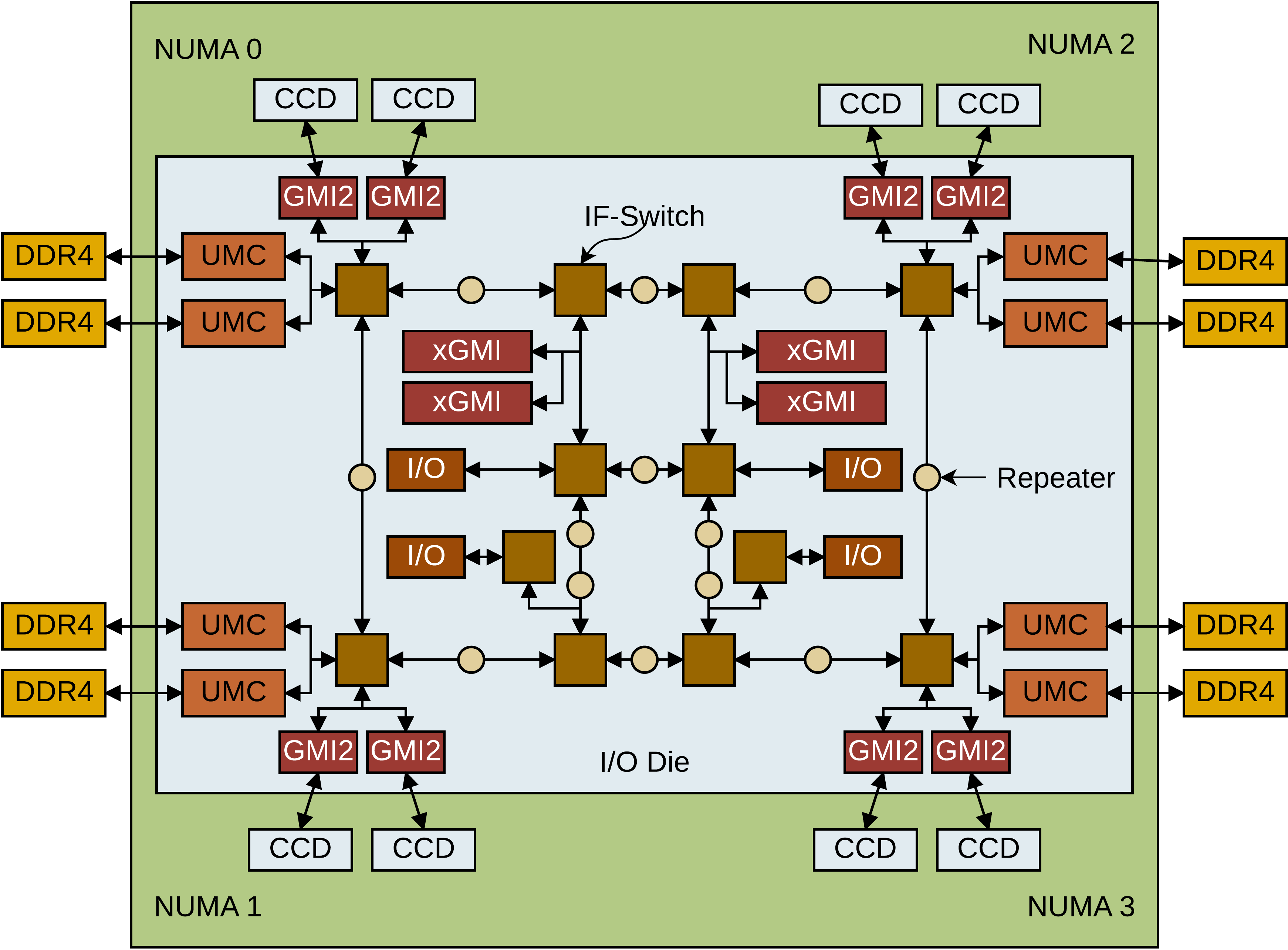}
    \caption{\label{fig:rome-io-die}Layout of an AMD Rome processor with its central I/O-die, which connects CCDs, DRAM, and I/O via an Infinity Fabric (IF) network, Global Memory Interfaces (GMI), and Unified Memory Controllers (UMCs).~\cite{AMDRomeRR2020,Naffziger2021}}
	\Description{Visualization of an AMD Rome processor, centering the I/O-die. In each of the four corners of the I/O-die, an Infinity Fabric switch connects two CCDs (via GMI2), 2 Memory Channels (via UMC), and two IF-repeaters. While the vertical IF-repeaters are then connected to one of the other "corner" IF-switchess, which attach CCDs, the horizontal one forward to switches that connect the external connections like xGMI, PCIe and other I/O.}
\vspace{5mm}
\end{figure}

Rome processors use two different types of dies that are combined on one package~\cite{Suggs2020, Naffziger2020Paper}.
Up to eight Core Complex Dies (CCD) are connected to the I/O-die via AMD's Infinity Fabric (IF).
The I/O-die (\figref{rome-io-die}) and its Infinity Fabric (IF) grid connect the CCDs among each other and to external components, including the 128 PCIe Gen~4 lanes~\cite[Figure 21]{AMDRomeRR2020},~\cite{Naffziger2021} and main memory.
IF-switches are used to route data through the I/O-die, which induce a latency of at least 2 Fabric Clock (FCLK) cycles at the nominal IF frequency of \unit{1467}{\mega\hertz}~\cite{Naffziger2021}.
Additionally, IF-repeaters are used, which cause a 1 FCLK cycle latency~\cite{Naffziger2021}.

\begin{figure}[p]
    \centering
    \includegraphics[width=.7\columnwidth]{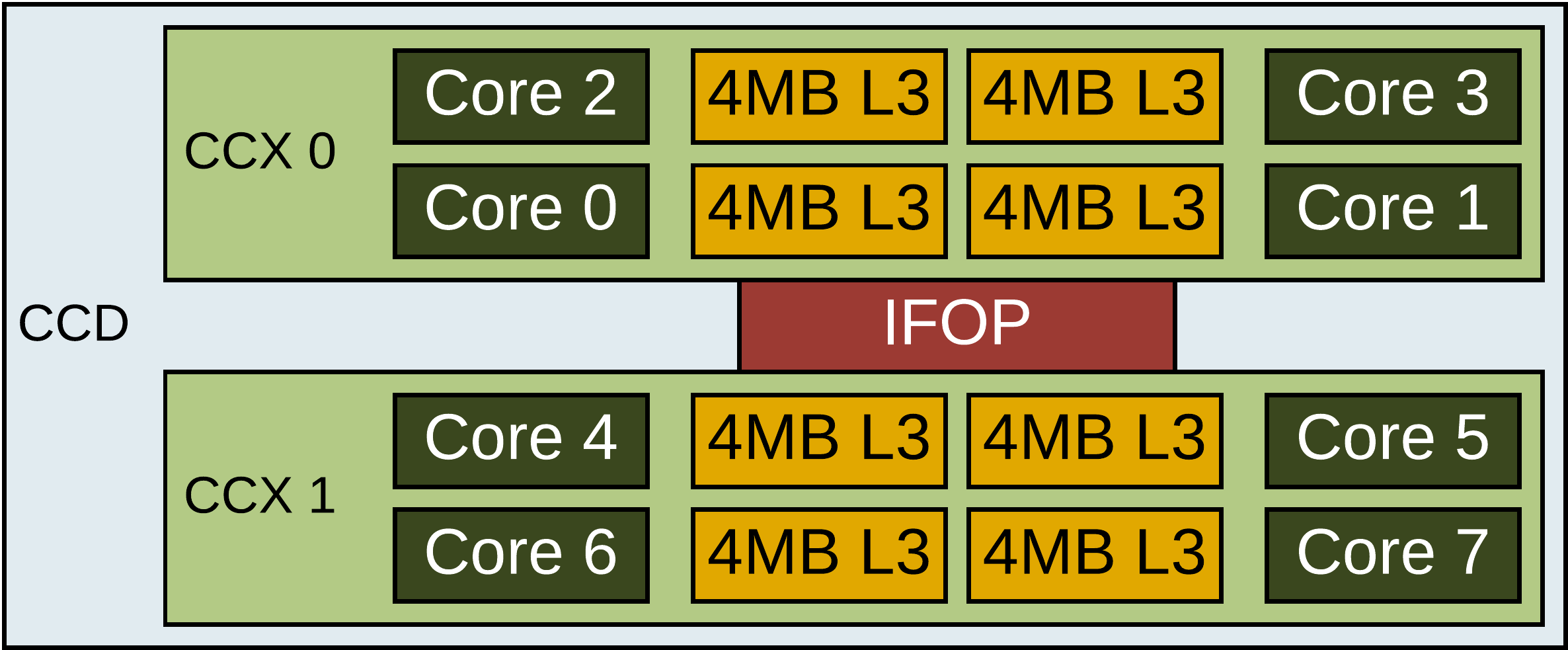}
    \caption{\label{fig:zen2-ccd}Layout of an AMD Rome Core Complex Die (CCD) hosting two Core Complexes (CCX) and the Infinity Fabric On-Package (IFOP).~\cite{Suggs2020, Naffziger2020Paper}}
    \Description{Layout of an AMD Rome CCD, which hosts two separated CCX and connects via IFOP to external resources (I/O die). Each CCX hosts four cores, where each core holds a Level-3 cache slice that is shared with other cores of the same CCX.}
\vspace{5mm}
\end{figure}

\begin{figure}[p]
    \includegraphics[width=\columnwidth]{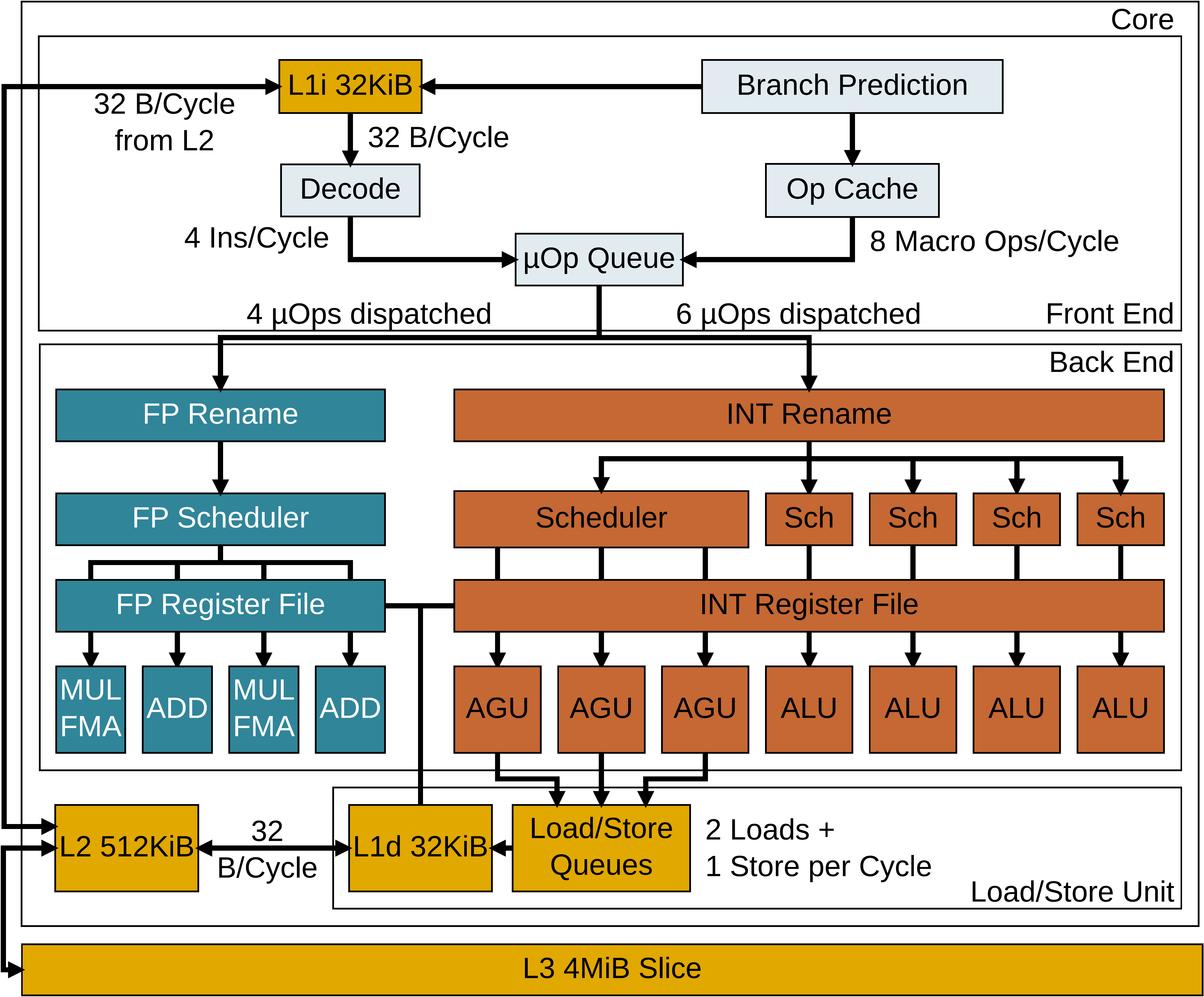}
    \caption{\label{fig:zen2-core}Layout of an AMD Zen 2 core (see~\cite{AMDOptimization2020,Suggs2020}).}
    \Description{Layout of an AMD Rome core, connecting to the Level-3 cache slice. Instructions can be loaded to the 32 KiB Level-1 instruction cache from the 512 KiB Level-2 cache with 32 B/cycle. They can then be fetched (32 B/cycle), decoded (4 instructions/cycle) and sent to the µOp queue, alternatively up to 8 Macro Ops can be fetched from the OpCache. Based on their type, operations are then sent to the separate integer and float parts of the core and processed by the following pipelines 2*FADD+2*FMUL/FMA (float), 3*AGU+4*ALU (integer). The AGUs can perform 2 loads and 1 store per cycle from/to the 32 KiB Level-1 data cache, which can access the Level-2 cache with 32 B/cycle}
\end{figure}

Each CCD includes two Core Complexes (CCX), which consist of up to four Zen 2 cores each (see \figref{zen2-ccd}), resulting in up to 64 cores per processor.
While L1 and L2 caches are per core (see \figref{zen2-core}), all cores within a CCX share a common \unit{16}{\mebi\byte} L3 cache.
The Infinity Fabric on Package (IFOP) interface on each CCD connects the two CCX to the I/O-die, but not to each other~\cite{AMDRomeRR2020,Naffziger2020Paper}.


\subsection{Memory Architecture Details}\label{sec:zen2-core}

Burd et al. published in~\cite{BurdZeppelin} that Zen uses a MDOEFSI (\emph{Modified, Dirty, Owned, Exclusive, Forward, Shared, Invalid}) protocol, without naming details.
They also state that Zen's Infinity Fabric is based on an enhanced coherent HyperTransport protocol which has been used by AMD before (compare~\cite{MolkaDiss}).
It can be assumed that these protocols are used in Zen 2 as well.

The layout of a Rome processors indicates the presence of four non-uniform memory access (NUMA) nodes.
BIOS settings can be used to expose NUMA nodes to the operating system (OS), one, two and four nodes can be configured~\cite[Section 2.5]{Kashyap2020} where \mycite{[m]emory is interleaved across the [...] memory channels in each NUMA domain}.
Also, \mycite{each server can be configured [...] with an additional option to configure L3 cache as NUMA nodes}, enabling up to 16 NUMA nodes per processor.
Due to the positions of the socket-to-socket Global Memory Interconnect (xGMI) interfaces, remote socket access latencies will depend on the relative position of the communication partners.
The architecture supports up to two sockets.

Each of the cores of a CCX hold one slice of the victim L3 cache~\cite{AMDOptimization2020}.
The L3 contains evicted cache lines from the L2 caches and valid copies of cache lines shared by multiple cores.
Also, data transfers and cache coherency between L2 caches on the CCX are managed by the shadow tags in the L3.
The L3 cache slices use a common L3 frequency, which is generally defined by the highest core frequency of all cores within the CCX~\cite{romeee}.

AMD Zen 2 cores use three Address Generation Units (AGU) in the integer execution unit to perform up to two \num{256}-bit loads and one \num{256}-bit store per cycle~\cite{AMDOptimization2020, Suggs2020}.
Compared to first generation Zen processors, these have been widened to match the increased width of SIMD execution.
Zen~2 floating point units are able to process \num{256}-bit Advanced Vector Extensions (AVX) instructions in one cycle~\cite{AMDOptimization2020, Suggs2020}, even though this could limit core frequencies in some cases~\cite{romeee}.
Load and store queues attached to the AGUs have 44 and 48 entries, respectively, and load data from a \unit{32}{\kibi\byte} L1d
cache~\cite{AMDOptimization2020, Suggs2020}.
Instructions are loaded at \num{32}\,B/cycle from a \unit{32}{\kibi\byte} L1i cache and buffered in a \unit{4096}{entry} op cache~\cite{Suggs2020}.
The \unit{512}{\kibi\byte} L2 cache is inclusive of both L1 caches~\cite{AMDOptimization2020, Suggs2020}.
AMD uses hardware prefetchers for the L1d, L1i, and the L2 cache to avoid stalls due to cache misses~\cite[Section 2.1]{AMDOptimization2020}.
L1 and L2 prefetchers can be deactivated via the BIOS, or, according to our findings, by disabling bit 0 of MSR \texttt{0xc001102b} and enabling bit 16 of MSR \texttt{0xc0011022} for each core.

\section{The Intel Cascade Lake SP Architecture}\label{sec:CLX-arch}

\subsection{General Concept}\label{sec:CLX-concept}
The CLX processor architecture succeeds the SKX architecture. Both are using a \unit{14}{\nano \meter} process and a monolithic design~\cite{ArafaCLX2019} and are available with up to 28 cores per processor.
As described in~\cite{ArafaCLX2019} and~\cite{Alappat2020}, changes between these two generations are minimal, including
higher core frequencies, support for Optane DC, faster DRAM, and additional instructions.
This is confirmed by Alappat et al., who note in~\cite[Section 2]{Alappat2020} that SKX and CLX behave identically in memory and floating point benchmarks.

A grid of horizontal and vertical fabric lanes connects all parts of the processor~\cite{ArafaCLX2019, Schoene2019, Tam2018} (see \figref{CLX-layout}).
Transfers within the 2D-mesh are always routed along the vertical axis, followed by the horizontal direction~\cite{Tam2018}.
All uncore components operate on a common frequency which is managed by a hardware control loop from a range of frequencies~\cite{Schoene2019, Tam2018}.
The number of PCIe ports, and thus lanes, depends on the number of cores, since entire columns or single cores of the grid may be deactivated/removed~\cite{IntelCLXBrief, McCalpinSKXTiles}.
This is a significant difference to AMD EPYC processors, where I/O interfaces are identical for all core counts.

\subsection{Memory Architecture Details}\label{sec:CLX-core}

\begin{figure}[b]
    \centering
    \includegraphics[width=\columnwidth]{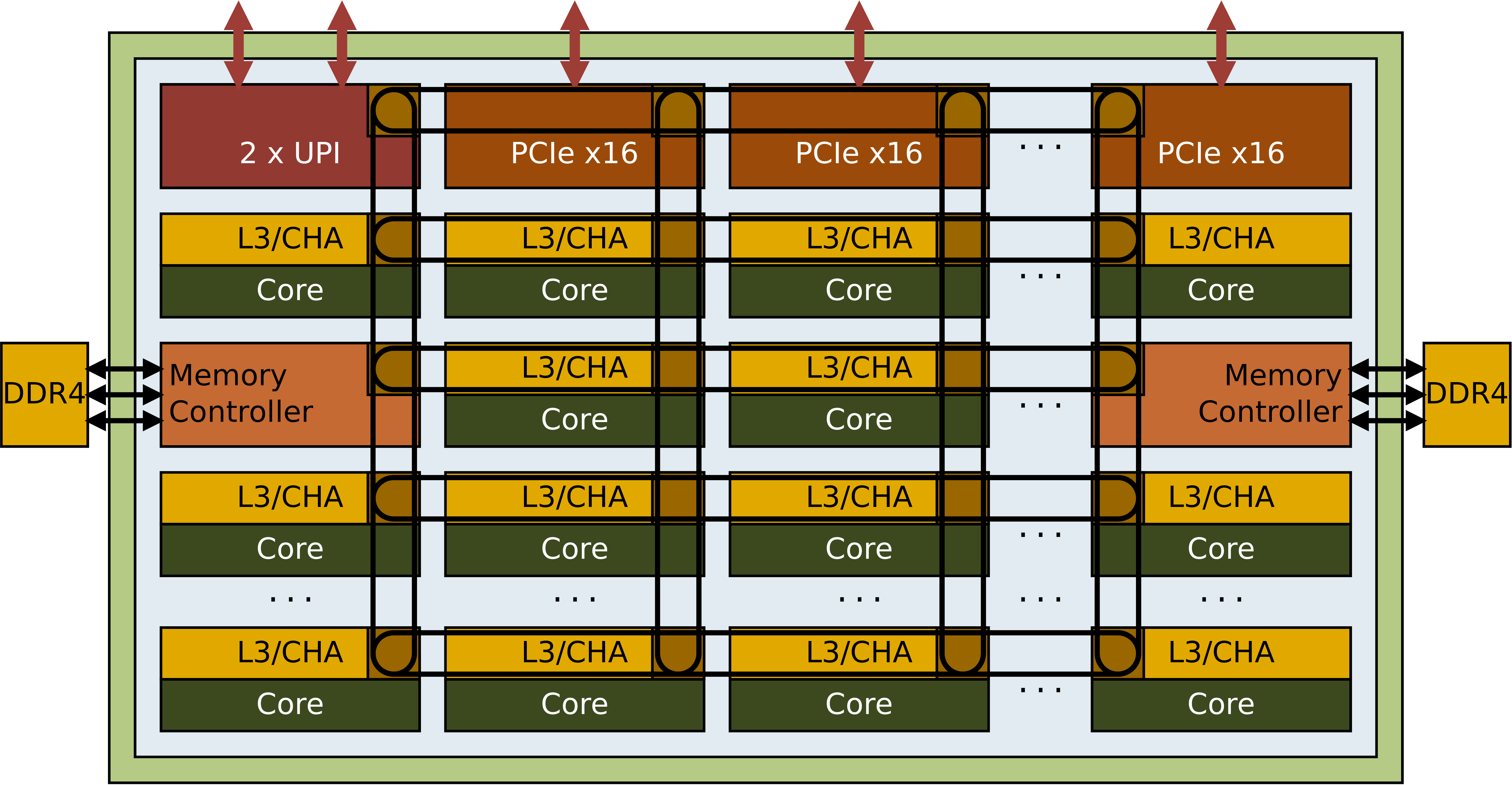}
    \caption{\label{fig:CLX-layout}Layout of an Intel Cascade Lake SP processor, where a 2D-mesh connects core tiles, memory controllers and I/O~\cite{Schoene2019,McCalpinSKXTiles}.}
	\Description{Mesh-layout of an Intel Cascade Lake processor with I/O connections (UPI, PCIe, ...) on the top side and core tiles below. in the second row of core tiles the most left and most right core tile is replaced with an integrated memory controller, which attaches three memory channels. A core tile holds an Level-3 cache slice, the CHA and a core.}
\end{figure}
\begin{figure}[b]
    \centering
    \includegraphics[width=\columnwidth]{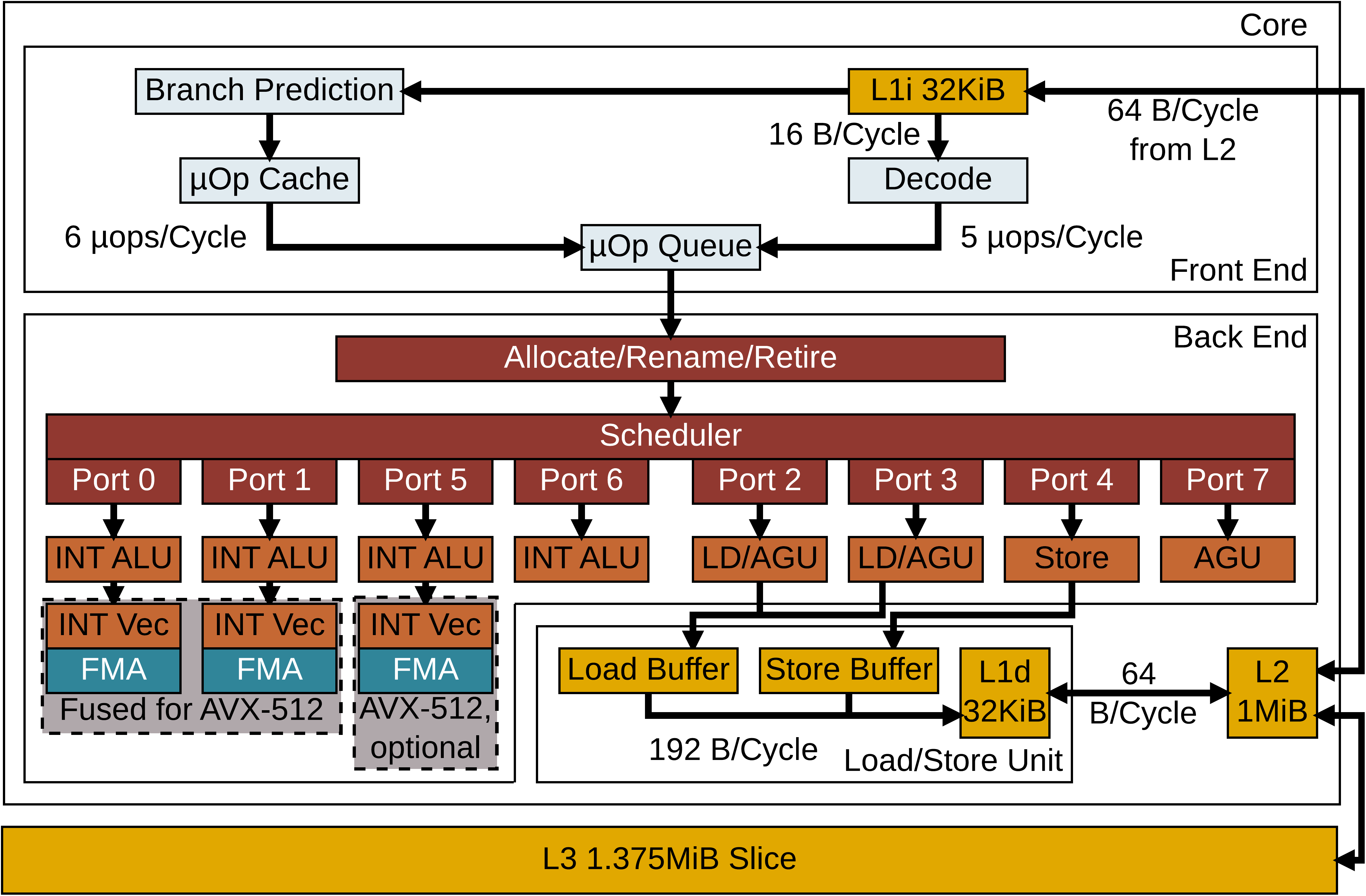}
    \caption{\label{fig:CLX-EU}Layout of an Intel Cascade Lake SP core} 
    \Description{Layout of an Intel Cascade Lake core, connecting to the 1.375 MiB Level-3 cache slice. Instructions can be loaded to the 32 KiB Level-1 instruction cache from the 512 KiB Level-2 cache with 64 B/cycle. They can then be fetched (16 B/cycle), decoded (5 µOps/cycle) and sent to the µOp queue, alternatively up to 6 µOps can be fetched from the µOp cache.The collected µOps are processed by the different ports, including AGUs for memory accesses. The AGUs can perform 2 64-B loads and 1 64-B store per cycle from/to the 32 KiB Level-1 data cache, which can access the Level-2 cache with 64 B/cycle}
\end{figure}
Intel uses a unified execution unit design for CLX, unlike AMD (\secref{zen2-core})~\cite{IntelOptimization2020,AgnerMicroarchitecture}.
Four AGUs are available: two for \num{512}-bit loads, one for \num{512}-bit stores and one for address generation only~\cite[Table 11.1]{AgnerMicroarchitecture}.
Port 6 is reserved for integer/logic-only instructions.
The read  buffers have 72 entries, 56 entries are available in the write buffers~\cite[Section 11.9]{AgnerMicroarchitecture}.
Both, integer and SIMD floating point instructions, can be executed by ports 0, 1 and 5.
Ports 0 and 1 can be fused for the execution of AVX-512 vector instructions.
On some processors, port 5 has as a second AVX-512 unit, Intel lists the number of available AVX-512 units in~\cite[Page 10]{IntelOptimization2020}.
When executing AVX(-512) instructions, cores  use dedicated frequencies in order to prevent thermal damages to the processor.
These depend on the type of instruction and the number of cores that are running that instruction~\cite{IntelCLXSpecUpdate}.
Sch\"{o}ne et al. describe side effects for transitions between normal and AVX-512 frequencies for SKX in~\cite{Schoene2019}.

CLX processors have a L1i and L1d~cache of \unit{32}{\kibi \byte} (\figref{CLX-EU}), equal to Zen 2.
Both caches load data at \num{32}\,B/cycle.
A 1536 entry op cache buffers instructions~\cite[Table 11.12]{AgnerMicroarchitecture}.
The L2~cache is inclusive of the L1 caches and at \unit{1}{\mebi\byte} twice as large as the Zen 2 L2~\cite{Tam2018}.
The L1 and L2 caches utilize hardware prefetchers to reduce the impact of cache misses~\cite[Section E.3]{IntelOptimization2020}.

As indicated in \figref{CLX-layout}, a core-tile holds not only a core, but also a slice of \unit{1.375}{\mebi \byte} L3 cache and the Caching and Home Agent (CHA)~\cite{Tam2018}.
The CHA maps accessed addresses outside its own caches and provides the necessary routing information through the mesh~\cite{MulnixSLX2017}.
CLX has a non-inclusive L3 cache, which \mycite{may appear as a victim cache}, \mycite{depending on the access pattern, size of the code and [accessed] data, and sharing behavior between cores}~\cite[Section 2.2.1.2]{IntelOptimization2020}~\cite{KumarSLX}.
Generations prior to SKX used smaller L2 and larger per-core L3 caches, the latter being inclusive.
According to~\cite[Section \textquotedblleft Cache Hierarchy Changes\textquotedblright]{MulnixSLX2017}, the new cache layout was chosen to reduce overall latencies, as more requests may be served from the larger L2. 
The non-inclusive L3 leads to a higher effective total cache size, since it does not necessarily hold data that is present in other cache levels.
Although the per-core L3 cache size is higher for Zen 2 (\unit{4}{\mebi \byte} over \unit{1.375}{\mebi \byte}), Intel's 2D mesh interconnect allows all cores to access all L3 slices with a worst-case overhead of 18 uncore cycles\footnote{9-hop distance for a 28 core configuration}~\cite{MulnixSLX2017}.
However, Intel implements a suboptimal slice hash mechanism, which can lead to an imbalance in using the available L3-slices~\cite{McCalpinCacheHash}.

SKX and CLX processors have two memory controllers with three memory channels each~\cite{Schoene2019}.
They can be split into two NUMA nodes, referred to as Sub-NUMA  Clusters (SNC), by assigning each half of the cores to one of the two memory controllers as well keeping the L3 cache slices local to the core's respective SNC~\cite{MulnixSLX2017}.

SKX processors implement the MESIF coherency protocol with the states \emph{Modified}, \emph{Exclusive}, \emph{Shared}, \emph{Invalid}, \emph{Forwarded}~\cite[Section 2.2.4]{IntelCASUncorePerf2017}.
The directory protocol names two additional states: \emph{Any} and \emph{L}\footnote{as defined in description, or \emph{P} as defined in unit mask name}~\cite[Table 2-179]{IntelCASUncorePerf2017}, which are not documented and for this reason not investigated in this paper.
We assume that CLX uses the same protocol.

\section{Test System \& Benchmarks}
\label{sec:test-system}
We performed our measurements on two dual socket servers with AMD EPYC 7702 and Intel Xeon Gold 6248 processors, respectively.
\tabref{test-system} lists relevant details and configurations.
Unless stated otherwise, we use nominal core frequencies.
The uncore frequency of the Intel system was set to its nominal frequency of \unit{2400}{\mega\hertz} with the \texttt{likwid-setFrequencies}\footnote{\url{https://github.com/RRZE-HPC/likwid/wiki/likwid-setFrequencies}} tool~\cite{likwid}.
While the uncore clock can be lowered if the Thermal Design Power (TDP) is reached~\cite{Schoene2019}, we designed our experiments in a way that such a behavior is not triggered.

\begin{table}[t]
    \caption{\label{tab:test-system}Overview on hardware and software of used test systems}
    {\setlength{\tabcolsep}{0.2em}
	\begin{tabular}{rrr}
	\hline
		Processor	&	2 $\times$ \textbf{AMD EPYC 7702} & 2 $\times$ \textbf{Intel Xeon Gold 6248} \\
		Cores		&	$2 \times 64$ & $2 \times 20$ \\
		Avail. freq.s	&	1.2, 1.5, \textbf{\unit{2.0}{\giga\hertz}}, Turbo & 1.2--\unit{2.5}{\giga\hertz}, Turbo \\
		L1 cache	& 	$128 \times (\unit{32}{\kibi\byte} + \unit{32}{\kibi\byte})$ & $40 \times (\unit{32}{\kibi\byte} + \unit{32}{\kibi\byte})$ \\
		L2 cache	&	$128 \times \unit{512}{\kibi\byte} = \unit{64}{\mebi\byte}$ & $40 \times \unit{1024}{\kibi\byte} = \unit{40}{\mebi\byte}$ \\
		L3 cache	&	$32 \times \unit{16}{\mebi\byte}=\unit{512}{\mebi\byte}$ & $2 \times \unit{24.75}{\mebi\byte}=\unit{49.5}{\mebi\byte}$ \\
		Server Model& GIGABYTE & HPE \\
 		/ Mainboard	& MZ62-HD0-00 & ProLiant DL360 Gen10 \\
 		Memory		&	Micron  & HPE P03052-091\\
                    & 18ASF4G72PDZ-3G2B2 & \\
                    &	$16\times\unit{32}{\gibi\byte}=\unit{512}{\gibi\byte}$ & $12\times\unit{32}{\gibi\byte}=\unit{384}{\gibi\byte}$\\\
                    &	\unit{1600}{\mega\hertz} & \unit{1467}{\mega\hertz} \\
		OS          &   CentOS 7.7.1908 (Core)  & Ubuntu 18.04\\
		Kernel      &   3.10.0 x86\_64 & 4.15 x86\_64\\
		NUMA-Setup	&	$2 \times 4 = 8$ (NPS4) &  $ 2 \times 2 = 4 $ (SNC)\\
\hline
		\multicolumn{3}{c}{\tiny{--- BenchIT ---}} \\
		Compiler    & GCC 10.2.0& GCC 7.5\\
\hline
		\multicolumn{3}{c}{\tiny{--- STREAM ---}} \\
		Compiler	& Intel 19.0.1 & Intel 19.0.5 \\ &\multicolumn{2}{c}{\multirow{2}{6.5cm}{\centering\texttt{-DSTATIC -DSTREAM\_ARRAY\_SIZE=800000000 -qopenmp -Ofast}}} \\ 
		\multirow{6}{1cm}{Comp. flags} & & \\
                    &   \texttt{-mcmodel=large}  & \texttt{-qopt-streaming-}\\  
					&   \texttt{-shared-intel} & \texttt{stores  always}\\ 
					& & \texttt{-march=cascadelake}\\ 
					& & \texttt{-xCORE-AVX512}\\ 
					& & \texttt{-qopt-zmm-usage=high} \\

		\hline
	\end{tabular}
	}
\end{table}

We use \emph{BenchIT}\footnote{\url{https://tu-dresden.de/zih/forschung/projekte/benchit/}}~\cite{JuckelandBenchIT}, with the \textit{x86-membench} extension~\cite{MolkaDiss} for latency and bandwidth analyses.
These benchmarks can be configured for a wide range of objectives by changing the configuration of the respective \texttt{PARAMETERS} file.
To analyze the impact of the cache coherence protocol, we use the benchmark \emph{memory\_latency}.

As described in~\cite[Section 3.5.1]{MolkaDiss}, the \emph{memory\_latency} benchmark uses pointer-chasing to determine latencies.
The size of the allocated buffer determines the memory level(s) to be analyzed.
The accessed addresses are created by an additional thread using a random number generator to minimize prefetcher-effects.
A coherence state control routine ensures that accessed cache lines have the requested coherence state on the requested core by running an additional thread on this core before the actual measurement.
In~\cite[Section 3.3]{MolkaDiss}, Molka describes how the different coherence states are created.
After measuring the start time (\texttt{rdtsc} serialized with \texttt{mfence} and \texttt{lfence} calls), the measurement thread accesses the memory addresses (loop with unrolled \texttt{mov (\%rbx), \%rbx}), and measuring the end time (serialized \texttt{rdtsc}).
The duration is then adjusted by the measurement overhead, which is determined with the same routine without the memory accesses.
\lstref{latency-benchmark} provides a high-level overview of the steps to measure the memory accesses.

\begin{lstlisting}[numbers=left, label=alg:latency-benchmark, basicstyle=\footnotesize, caption={\emph{memory\_latency} overview. Thread M is only required for coherence-states Shared, Forward, and Owned.{~\cite[Section 3]{MolkaDiss}} }]
pin threads 0,N(,M) via sched_setaffinity()
thread 0,N(,M): allocate memory via numa_set_membind()
thread 0: warm-up of TLB by touching memory
thread 0: signal thread 1 to prepare data
thread N(,M): touch data correctly [20, Sec. 3.3]
thread 0: wait for thread N(,M)
thread 0: access memory of thread N, measure latency
\end{lstlisting}

In order to achieve reproducible results, we flushed all cache levels before each run with the \texttt{BENCHIT\_KERNEL\_FLUSH\_\{L1|L2|L3\}=1} flag in the PARAMETERS file of the benchmark.
We use the default \unit{512}{\byte} alignment to avoid the re-use of cache lines, set with \texttt{BENCHIT\_KERNEL\_ALIGNMENT=512}, unless otherwise noted.

Cache and main memory bandwidth values are determined with the \emph{throughput} benchmark where we use different Streaming SIMD Extensions (SSE) and AVX instructions.
To ensure that our bandwidth results are truly limited by the achievable bandwidth, we configure the benchmark with the \texttt{BENCHIT\_KERNEL\_BURST\_LENGTH} parameter to use eight \unit{64}{\bit} xmm, 16 \unit{256}{\bit} ymm or 32 \unit{512}{\bit} zmm vector registers for SSE, AVX, and AVX-512 kernels, respectively.
In order to use the 32 zmm vector registers, additional compiler flags (\texttt{-mavx512f} in our case) have to be used.
We extended the benchmark with a range of AVX-512 kernels.
Transparent huge pages are used for both benchmarks to clearly distinguish memory levels, as advised in~\cite{MolkaDiss}.
Hardware prefetchers are enabled for all measurements on both systems, except for the latency benchmark on CLX.
We flush all cache levels before each run in order to achieve reproducible results and to avoid the impact of prefetchers.
We set \texttt{/sys/kernel/mm/\ transparent\_hugepage/enabled} to \texttt{always} to clearly distinguish memory levels.

We configure BenchIT to use four dataset sizes per memory level.
For each of these values, BenchIT reports the minimum for latencies or maximum for bandwidths of three internal measurements to filter out outliers from external influences.
We repeat each experiment ten times resulting in a total of 120 values (10$\times$4$\times$3) per reported data point, combined using minimum and maximum.
As latencies for remote L1 accesses were noisy, we report the medians for these measurements, which coincide with the modes of the samples.

Additional measurements are performed with \emph{STREAM}, as it provides more realistic memory access patterns with loads and stores.
We use non-temporal stores in order to achieve a higher bandwidth closer to theoretical limits.
\emph{STREAM} repeats measurements ten times internally and reports the highest bandwidth.
In addition, we run it ten times and use the maximum of the reported values.

\section{Local Memory Access}
\label{sec:local}

\subsection{Latencies}
\label{sec:local-latency}
\begin{figure}[t]
    \centering
    \includegraphics[width=\columnwidth]{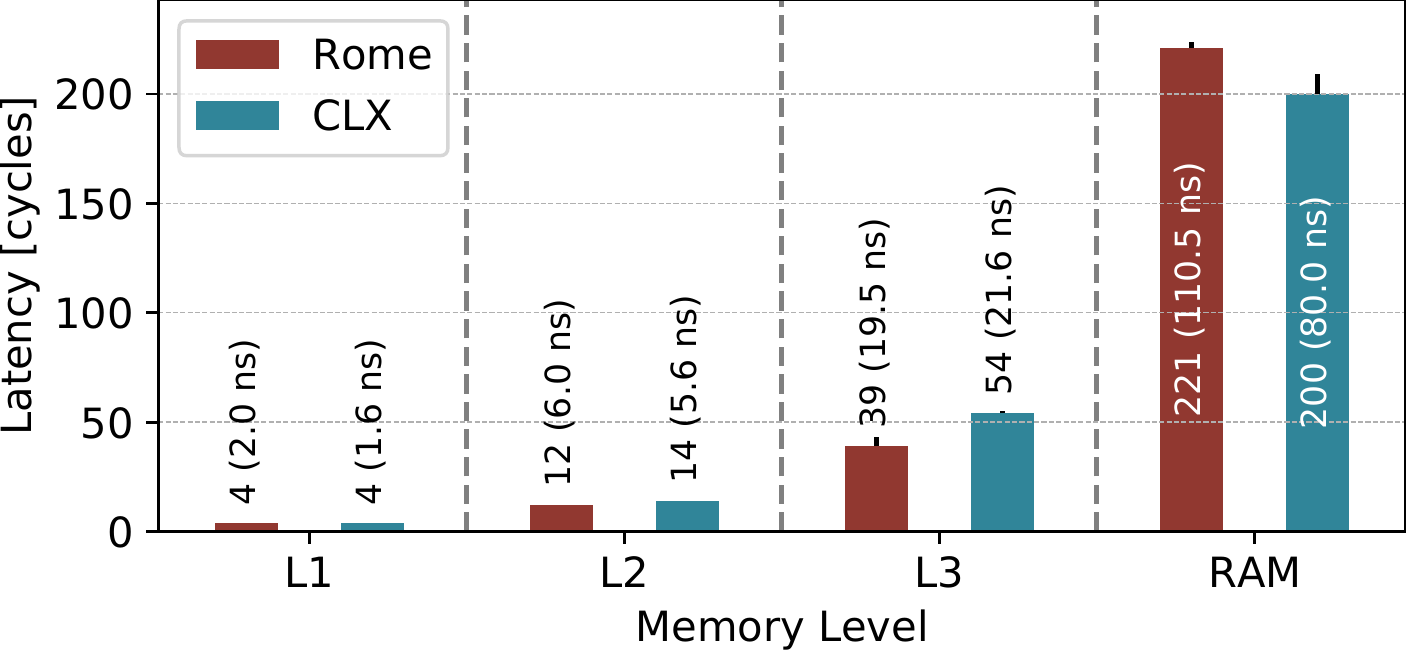}
    \caption{\label{fig:local-accesses}Local memory access latencies for AMD EPYC 7702 (Rome) and Intel Xeon Gold 6248 (CLX), error bars indicate maxima.}
	\Description{Bar chart for local memory latencies for AMD EPYC 7702 (Rome) and Intel Xeon Gold 6248 (CLX). Results are also mentioned in the text.}
\end{figure}

In the simplest case, processor cores access data in their own memory hierarchy without influences from additional cores and NUMA domains.
We compare the local latencies of the two architectures, using the BenchIT \textit{memory\_latency} benchmark, in \figref{local-accesses}.
To focus on the efficiency of the architecture rather than the applied core frequency, we mainly present results in cycles and not in ns.
The L1d caches can be accessed with \num{4}\,cycles on both architectures (\num{2} and \unit{1.6}{\nano\second} for the Rome and CLX architecture, respectively) . 
L2 cache read latencies for Rome and CLX are \num{12}\,cycles (\unit{6}{\nano\second}) and \num{14}\,cycles (\unit{5.6}{\nano\second}), respectively.
However, the Intel CLX platform provides twice as much L2 cache per core.
This reduces the probability for L2 cache misses and even less efficient L3 cache accesses.
For the locally accessible L3 cache\footnote{We set \texttt{BENCHIT\_KERNEL\_ALIGNMENT=64} for L3 latency tests on the Rome system to use all L3 slices equally.}, Intel provides more capacity available for a single core, but at higher access latencies (\num{54}\,cycles (\unit{21.6}{\nano\second}) vs \num{39}\,cycles (\unit{19.5}{\nano\second})).
While single-threaded applications will benefit from larger local L2 and L3 caches on Intel platforms, the \emph{total} size of L2 and L3 caches is higher for AMD processors, which is beneficial for parallel applications.
Moreover, for equally clocked processors, the AMD system provides lower latencies to caches, which further reduces memory stall cycles.

Main memory latencies cannot be easily compared between processors, since the specifications of the installed memory DIMMs also contribute to their access time.
Moreover, adapting I/O p-states~\cite{romeee} and uncore frequencies~\cite{Schoene2019} lead to more uncertainty.
We measure a latency of about \num{220}\,cycles (\unit{110}{\nano\second}) for the Rome system, and \num{200}\,cycles (\unit{80}{\nano\second}) for CLX.
The higher RAM latencies for Rome may be attributed to the data path routing via the I/O-die, whereas the memory controllers of the CLX processor are integrated into the fabric mesh.

\subsection{Bandwidth}
\label{sec:local-bandwidth}
HPC applications often access main memory with a predictable pattern, which allows the compiler (via instruction re-ordering), the out-of-order engine, and hardware prefetchers to hide occurring latencies and thereby improve the effective memory bandwidth.
However, bandwidth is also limited by the width of datapaths and concurrent access to shared resources, i.e., L3 cache and DRAM.

We use the BenchIT \emph{throughput} kernel for our analysis, which only performs loads.
This kernel relies on a streaming access to memory using SIMD extensions to benefit from wider load instructions.
We extended the kernel so that 512-bit wide accesses can be measured.
Intel SKX and CLX processors have dedicated AVX(-512) frequency ranges with dedicated nominal frequencies, which directly influence these measurements.
To avoid an unwanted and unpredictable change of frequencies, we pinned core frequencies to \unit{1.6}{\giga\hertz} -- the nominal core frequency for AVX-512 for the Intel Xeon Gold 6248 processor~\cite[Figure 3]{IntelCLXSpecUpdate}.
Since AMD did not openly publish dedicated AVX frequencies, we use the nominal frequency (\unit{2.0}{\giga\hertz}) for the Rome system, even though highly demanding workloads could lead to throttling~\cite[Section V.E]{romeee}.

\begin{figure}[b]
    \centering
    \includegraphics[width=\columnwidth]{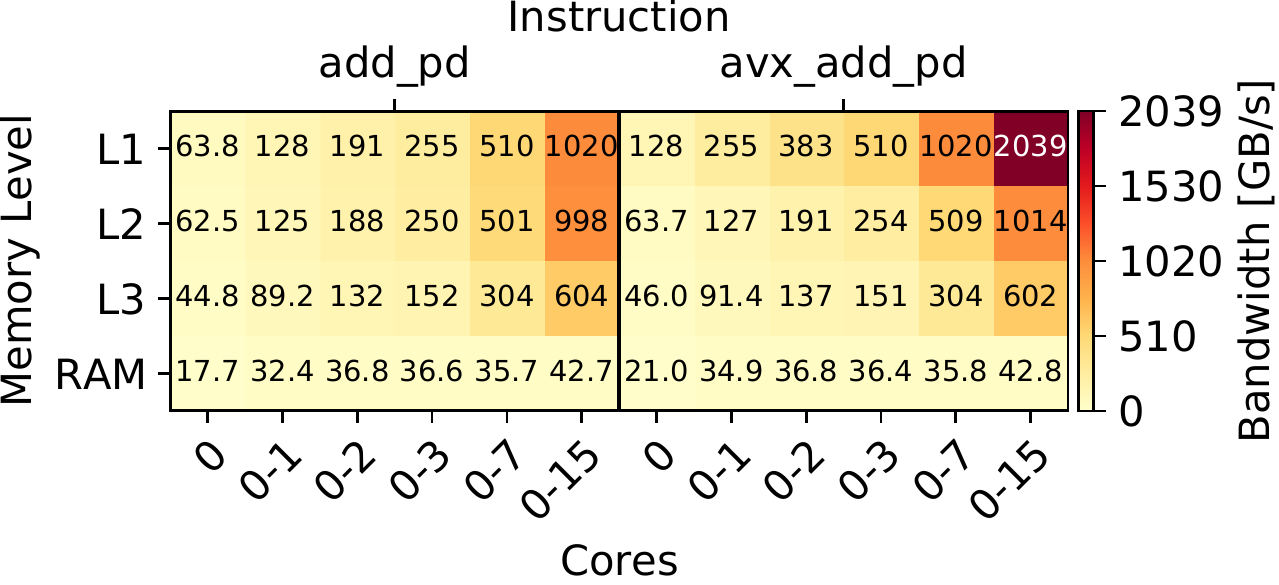}
    \caption{\label{fig:rome-throughput-local}AMD EPYC 7702 -- bandwidth at reference frequency (\unit{2}{\giga\hertz}) for one NUMA node.}
	\Description{Heat map for AMD EPYC 7702 (Rome) memory bandwidth to different memory levels (L1, L2, L3, main memory) and various core counts (1|2|3|4 cores on one CCX, 8 cores on one CCD, 16 cores on 2 CCDs). All measurements are performed on a single NUMA node. Measurements are taken for SSE and AVX read accesses.}
\end{figure}

\figref{rome-throughput-local} shows bandwidth results for memory reads on the AMD Rome system. 
The AMD cores reach the theoretical L1 bandwidth of \unit{128}{\giga\byte\per\second} (\num{64}\,B/cycle) if two \num{256}-bit floating point pipes are utilized. 
The bandwidth for the L2 cache falls slightly short of their theoretical maximum, with measured \unit{63.7}{\giga\byte\per\second} (\num{31.4}\,B/cycle) instead of \unit{64}{\giga\byte\per\second}.
A single core can read from the L3 with up to \unit{46}{\giga\byte\per\second} (\num{23}\,B/cycle) when using AVX.
According to our measurements, bandwidth for the L3 does not scale linearly for an entire CCX.
With four cores the achieved bandwidth is only \unit{151}{\giga\byte\per\second} (\num{18.9}\,B/cycle/core).
Also, RAM bandwidth for one CCD is saturated by using 3 cores of a single CCX in these measurements.
While adding accesses from the second CCX on one CCD (cores 0-7) does not help in gaining more performance, using the second CCD of a NUMA node (cores 0-15) increases the measured RAM bandwidth slightly.
The results for SSE (\texttt{add\_pd} instruction) and AVX do not differ significantly, except for L1.

\begin{figure}[t]
    \centering
    \includegraphics[width=\columnwidth]{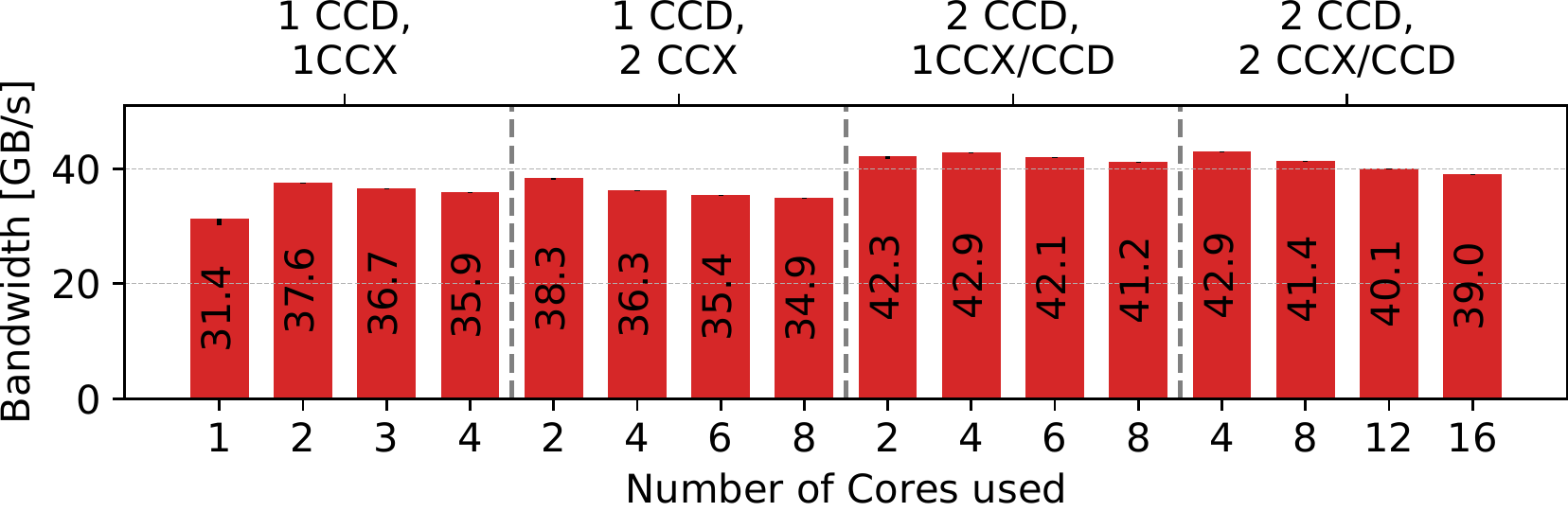}
    \caption{\label{fig:rome-stream-node}AMD EPYC 7702 -- \emph{STREAM} Triad: bandwidth measured on a single NUMA node with two CCDs and two CCX per CCD, error bars indicate minima.}
	\Description{Bar chart for STREAM Triad bandwidth when accessing memory of a single NUMA node with different cores of that NUMA node. Measurements for: (1,2,3,4) cores on one CCX, (2,4,6,8) cores on two CCX of the same CCD, (2,4,6,8) cores on two CCX of two CCDs, (4,8,12,16) cores on four CCX of both CCDs.}
\end{figure}

Since an increase of used cores beyond three per CCX does not increase bandwidth, we evaluate further bandwidth saturation points:
\figref{rome-stream-node} shows bandwidth measurements on a single NUMA node with two CCDs hosting two CCX each.
We use \emph{STREAM} for this analysis to achieve higher bandwidths for low core counts (due to non-temporal stores).
We scaled the number of cores per CCX and thus per CCD of a single NUMA node.
The highest bandwidth of a single NUMA node (\unit{42.9}{\giga\byte\per\second}) was measured in configurations with two participating CCDs and two cores per CCD. 
Since each NUMA node has its own set of memory channels, this bandwidth scales to \textasciitilde\unit{171}{\giga\byte\per\second} for the entire socket, as dedicated measurements show (not depicted).
Even with only one core per CCD, the measured bandwidth is only slightly lower at \unit{42.3}{\giga\byte\per\second}.
These numbers indicate that a small number of cores per CCX should be used for bandwidth-limited workloads.
Our measurements imply that users with a demand for a high bandwidth may benefit from processors with two CCDs per NUMA node, even if the higher core counts might not be necessary with respect to compute performance.

\begin{figure}[t]
    \centering
    \includegraphics[width=\columnwidth]{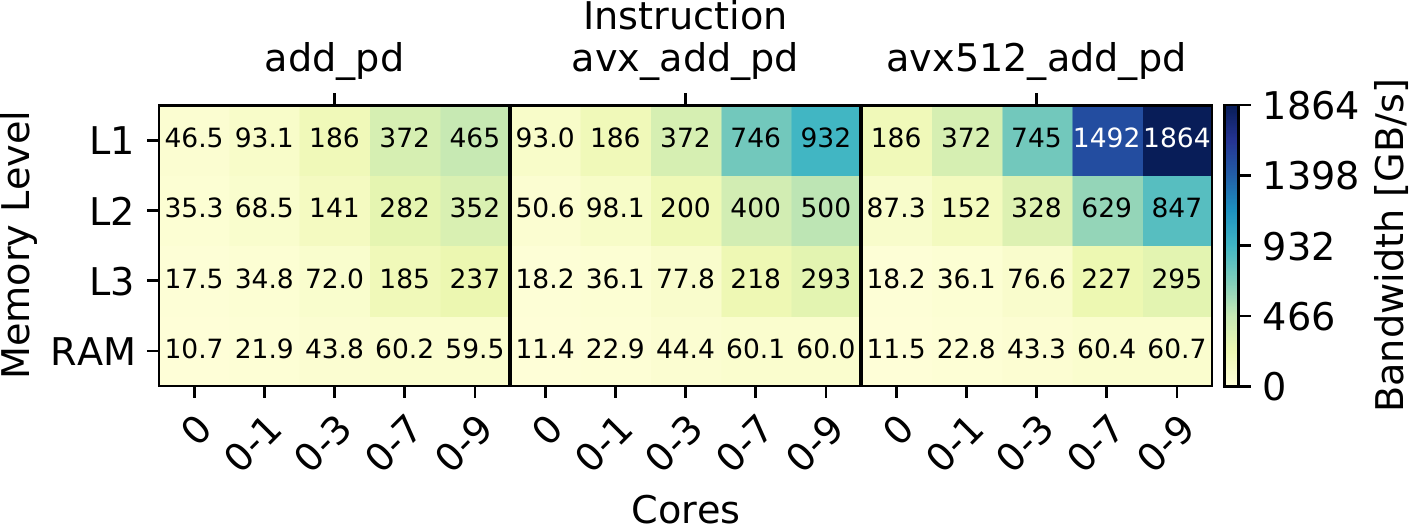}
    \caption{\label{fig:CLX-throughput}Intel Xeon Gold 6248 -- bandwidth at nominal AVX-512 frequency (\unit{1.6}{\giga\hertz}) for one SNC}
	\Description{Heat map for Intel Xeon Gold 6248 memory bandwidth to different memory levels (L1, L2, L3, main memory) and various core counts (1|2|4|8|10 cores on one SNC). Measurements are taken for SSE, AVX, and AVX-512 read accesses.}
\end{figure}

\figref{CLX-throughput} shows bandwidth results for reading memory accesses on the Intel Cascade Lake system.
The results do not reach the theoretical bandwidth of the L1d caches of \num{128}\,B/cycle when two \num{512}-bit floating point units are used.
Instead, we measure \num{116.25}\,B/cycle for AVX-512 instructions.
This is lower than the sustained bandwidths listed in~\cite[Table 2-6]{IntelOptimization2020}, but in line with measurements in~\cite[Figure 3]{Alappat2020}.
Notably, the L2 cache bandwidth varies for different instructions despite being consistently lower than the respective L1 bandwidth.
The L3 bandwidth results are similar for all SIMD widths, achieving  \num{11.3}\,B/cycle.
Again, this value is lower than the sustained bandwidth of \num{15}\,B/cycle described in~\cite[Table 2-6]{IntelOptimization2020}.
The RAM bandwidth is barely affected by SIMD width.
On a SNC, the RAM bandwidth can be saturated with eight cores.

The prerequisites for our measurements on the Rome and CLX system are significantly different.
This needs to be taken into account when comparing the observed cache bandwidth of the two architectures.
Running the AMD processor at its base clock speed may be a best case scenario, whereas the AVX-512 frequency we chose for CLX is more of a worst case, in particular for SSE and AVX.
In real world workloads, both processors might be able to use turbo frequencies.
As demonstrated in~\cite{romeee}, the Rome processor might throttle to a lower, not publicly disclosed frequency.
Ultimately, the used frequencies, and thus achieved bandwidths, depend on the workload and its distribution across cores and the thermal conditions the systems operate in.
Nevertheless, a few conclusions may be drawn.
Notably, the measured values for the CLX processor are lower than the maximum bandwidth and in many cases even the sustained bandwidth that are defined in \cite[Table 2-6, Section 2.2.1.3]{IntelOptimization2020}.
Moreover, the L2 bandwidths depend on the used instruction.
In an ideal situation for this processor, the SSE and AVX bandwidths would be close to the maximum throughput of the execution units for both the L1 and L2 caches, as the bandwidth requirement for these instructions is well within the maximum cache bandwidths.

This ideal case can be observed for Rome processors.
Here, the L2 bandwidth for SSE instructions is very close to the L1 bandwidth as well as to the L2 bandwidth for AVX instructions and the theoretical maximum.

The main memory bandwidth of a single NUMA node of a CLX processor is higher than on Rome, as CLX has three memory channels per NUMA node, Rome two.
However, the bandwidth of a NUMA node can be saturated with fewer cores on Rome than on CLX.
Additionally, the overall bandwidth for an entire socket for Rome is higher than for CLX, which is caused by Rome's eight memory channels compared to CLX' six.

\section{Intra-Socket Memory Latencies}
\label{sec:intra-socket}

While local memory accesses as described in the previous section are most common, thread migration, wrong data placements, and communication via shared memory lead to situations where a core accesses memory, which is not placed in its local memory hierarchy.
To measure latencies for such accesses, we again use the BenchIT \textit{memory\_latency} benchmark.
Threads are placed using the environment variables provided by x86-membench.
We distinguish accesses to cache lines placed in the coherency states \underline{M}odified, \underline{O}wned (on Rome) and \underline{E}xclusive, \underline{S}hared, \underline{I}nvalid, \underline{F}orward (on CLX).
We omit invalid cache lines from further analyses.
Since they are fetched from RAM at all times, the latencies are equal to those reported in \secref{local}.

\subsection{AMD EPYC Rome}
For Rome, we present Owned and Shared as well as Modified and Exclusive cache lines in combination as the observed performance pattern for each of these pairs is equal.
We performed measurements at nominal frequencies of \unit{2}{\giga\hertz}.

\begin{figure}[t]
 \centering
 \includegraphics[width=\columnwidth]{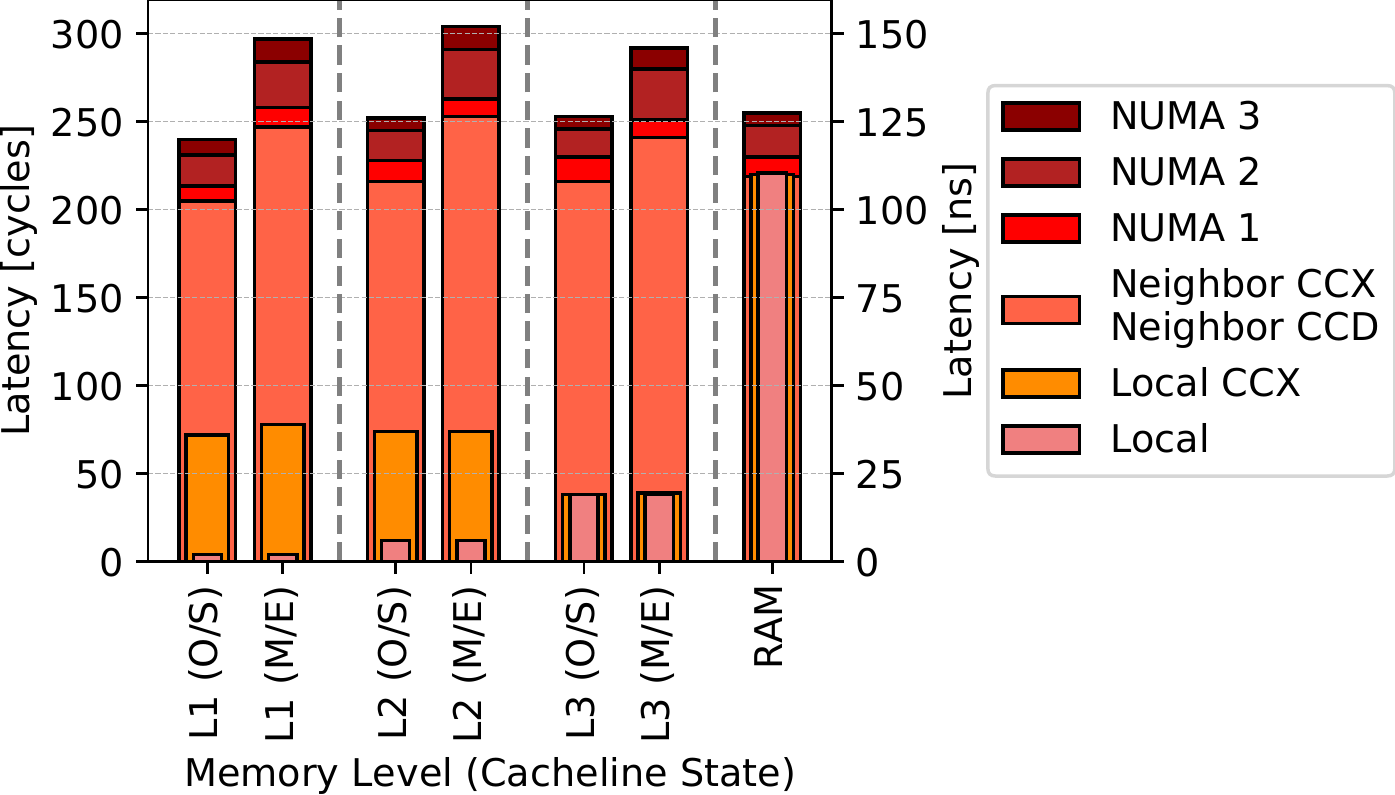}
 \caption{\label{fig:rome-socket-latencies}Memory read latencies for an AMD EPYC 7702 on one socket, measured from NUMA node 0 – 
     NUMA characteristics can be clearly observed.
     \emph{Owned} and \emph{Shared} as well as \emph{Modified} and \emph{Exclusive} states result in identical latencies. Refer \tabref{rome-1S-latencies} for precise values.}
 \Description{Stacked bar chart of latency measured when accessing data located in remote caches depending on core locations and MOESI state of the accessed cache lines.}
\end{figure}

\figref{rome-socket-latencies} summarizes the measurements for the Rome system:
When a core loads data from a different L1 cache within the same CCX, \emph{Owned} and \emph{Shared} cache lines can be served from the L2 cache within \num{72}--\num{74}\,cycles (\num{36}--\num{37}{\nano\second}).
This reflects the nature of L2 being inclusive of the L1 cache.
Modified and Exclusive data carry a slight penalty when accessed on a different L1 cache in the same CCX increasing latency to \num{78}\,cycles (\unit{39}{\nano\second}).
Data stored in the shared L3 cache of a CCX can be read within 
\num{39}\,cycles (\unit{19.5}{\nano\second}), which is about half the latency when accessing data in other L1 or L2 of other cores of a CXX.
According to~\cite{AMDOptimization2020}, the load-to-use latency of the L3 cache is \num{39}\,cycles.

Accesses to another CCX within the same NUMA node traverse through the I/O-die, since CCXs within a CCD are not directly interconnected.
At least a \num{200}\,cycles (\unit{100}{\nano\second}) latency is incurred when performing accesses to non-local L3 caches.

As shown in \secref{local}, reading from local RAM takes \num{220}\,cycles (\unit{110}{\nano\second}).
When reading memory from another NUMA node, the properties of the I/O-die can be observed. 
Accessing the RAM of the closest NUMA node requires additional  \textasciitilde\num{10}\,cycles (\unit{5}{\nano\second}).
This translates to two IF-switch and IF-repeater hops.
Assuming the IF operated at its nominal frequency of \unit{1467}{\mega\hertz}, this translates to \textasciitilde\num{3}--\num{4}\,FCLK cycles per switch and repeater combination.
A \textasciitilde\num{28}\,cycle (\unit{14}{\nano\second}) penalty is incurred for accesses to the second NUMA node, and \num{35}\,cycles (\unit{17.5}{\nano\second}) to the third NUMA node.
This implies that each IF-switch (\figref{rome-io-die}) in a path to another NUMA node incurs a \textasciitilde\num{2}--\unit{2.5}{\nano\second} latency per direction in our workload.

\subsection{AMD EPYC Rome - Intra-CCX}
\begin{figure}[b]
    \centering
    \includegraphics[width=\columnwidth]{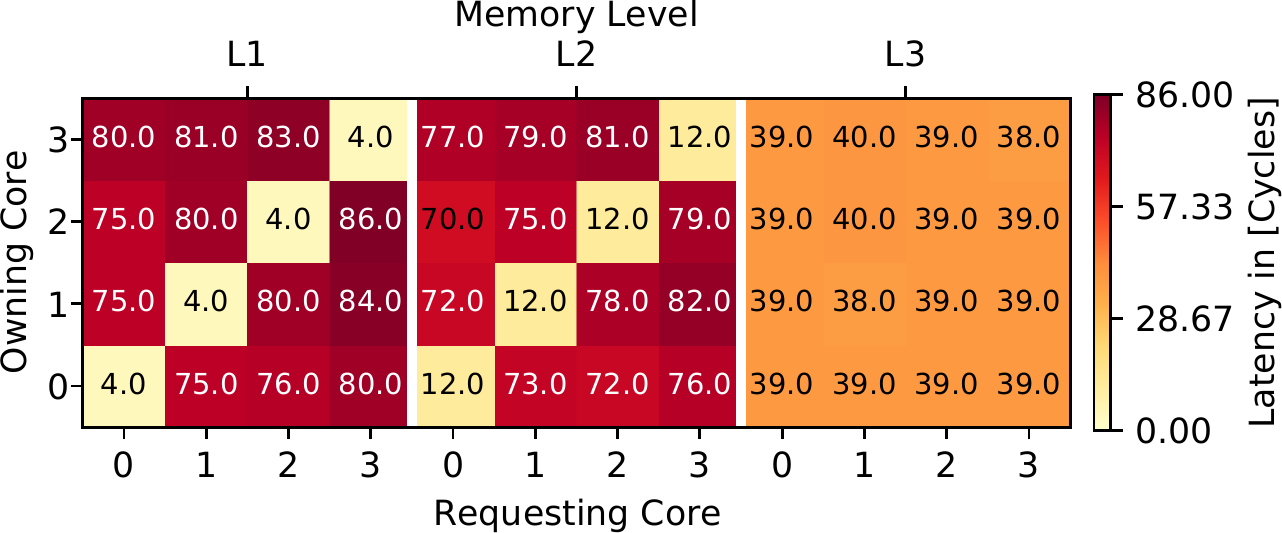}
     \caption{AMD EPYC 7702 -- CCX cache read latencies (\emph{Modified}): latencies for all communication patterns within a CCX. }
    \label{fig:rome-CCX-4x4-latencies}
	\Description{Three heat maps that visualize AMD EPYC 7702 (Rome) memory access latencies for Modified cache lines located in a remote cores L1, a remote cores L2, and the shared L3 within a CCX for different requesting/owning core combinations.}
\end{figure}

\figref{rome-CCX-4x4-latencies} shows latencies for all possible communication patterns within a CCX when \emph{Modified} cache lines are accessed.
L3 latencies seem uniform.
Accesses to other cores show a clear correlation between latencies and the relative position of the communicating cores within the CCX.
It is possible that these latencies are an effect of the \unit{512}{\byte} alignment, which we used for the L1 and L2 cache measurements.
With this alignment, the slice hash mechanism, which selects the L3 slice for a given cache line based on its address, will only use the L3-slice of core 0.
This gives us insight into the internal layout of a CCX:
If only the L3 slice of core 0 is accessed, the latency penalties for accesses will be caused by the distance between the used cores.
This can be observed when core 0 requests memory from other cores.
Cache lines held by cores 1 or 2 can be accessed within the same time. 
Requests to core 3 take additional time, i.e., \textasciitilde\num{5}--\num{7}\,cycles for the L1 and L2 caches.
A possible explanation would be that a ring interconnect is used between the cores/caches, where core 3 has a greater distance to core 0 than core 1 and 2.

\subsection{AMD EPYC Rome - Complex Request Flow}

\begin{figure}[t]
 \centering
 \includegraphics[width=0.5\columnwidth]{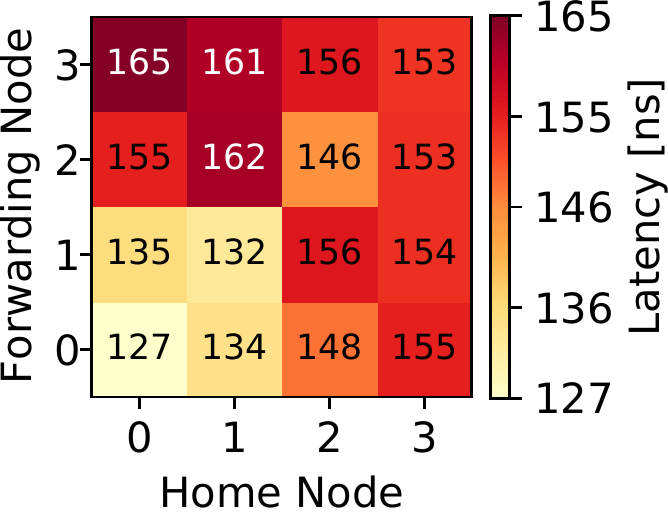}
 \caption{\label{fig:rome-request-flow} AMD EPYC 7702 -- L2 read latencies for complex cache accesses for \emph{Modified} cache lines: 
 NUMA node 0 requests memory which is allocated at the Home Node, but present in a \emph{Modified} state in the L2 cache of a core on the Forwarding Node.
 All potential scenarios for such accesses with node 0 as requesting node are
 evaluated.
 The diagonal from (0,0) to (3,3) indicates standard L2 latencies for direct access from node~0 with forwarding node being the home node.}
 \Description{Heat map visualizing latencies when accessing cache lines located in a remote L2 cache for different home nodes and forwarding nodes.}
\end{figure}

\figref{rome-request-flow} shows more complex cache request flows when accessing \emph{Modified} cache lines.
For example, a program is started on a core attached to NUMA node 2 and allocates its data there (\emph{home node}).
Later, this data is used and changed by a core at NUMA node 3 (\emph{forwarding node}) and holds the most recent copy of the memory in its cache.
Finally, the data is accessed by a core located at NUMA node 0 (\emph{requesting node}).
The data request would go from the requesting core to the memory controller of the home node and being forwarded to the core on the forwarding node, which replies to the request.
Please note that while three cores are involved in this scenario, the access pattern can be realized by a single thread, which is migrated by the operating system.
A core on another CCX is chosen for cases in which NUMA node~0 is also the forwarding node, as cache request would be fulfilled within the CCX otherwise.

\begin{figure*}[t]
    \centering
    \includegraphics[width=.95\textwidth]{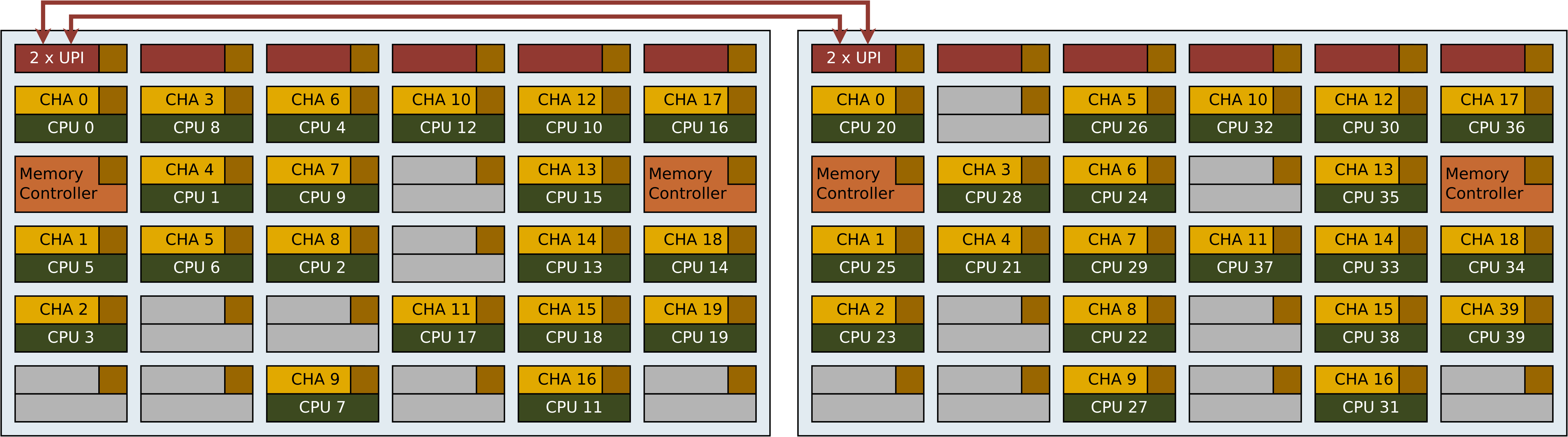}
    \caption{\label{fig:clx-tiles}Mapping of CPUs as listed by the OS to core tiles on CLX test system, deactivated core tiles are depicted in gray.
	\Description{Mapping of Core/CPU numbering to core tiles on the used Intel Cascade Lake test system. Each processor is composed of a 6x6 mesh, where the top row includes external communication. The only visualized node of these is in the top-left and represents the 2xUPI connection. between the processors. The rest of the meshes is as follows (X represents deactivated cores, M represents memory controllers):
	Processor 0: [[0,8,4,12,10,16],[M,1,9,X,15,M],[5,6,2,X,13,14],[3,X,X,17,18,19],[X,X,7,X,11,X]]
	Processor 1: [[20,X,26,32,30,36],[M,28,24,X,35,M],[25,21,29,37,33,34],[23,X,22,X,38,39],[X,X,27,X,31,X]]}
    }
\end{figure*}

The highest latency we measured in this scenario, \num{330}\,cycles (\unit{165}{\nano\second}), is higher than any latency for a native access in which home node and forwarding node are identical.
If the home as well as the forwarding node are different from each other and from the requesting node 0, the resulting latency is determined by the home node.
For instance, if NUMA node 1 is the home node, a latency of \num{322}--\num{324}\,cycles (\num{161}--\unit{162}{\nano\second}) is measured if either NUMA node 2 or 3 are forwarding nodes.
The same is true for other cases.
If, however, NUMA node 0 is both the requesting as well as the home node, the latency is determined by the choice of forwarding node.
Similarly, if NUMA node 0 is requesting and forwarding, the home node determines the latency.

Interestingly, the latencies are not symmetrical along the identity axis.
Instead, if NUMA node 0 is requesting and home node, the penalty of using a different forwarding node is higher than in cases where NUMA node 0 is requesting and forwarding node.
This is notable since in the first case the memory is already present at NUMA node 0 and must not necessarily be transferred to this node.
In cases where the forwarding and home node are not identical and not at NUMA node 0, the missing symmetry can be observed as well.
If NUMA node 1 is home node and NUMA nodes 2 or 3 are forwarding nodes, the latencies are almost identical.
However, if the roles of forwarding and home node are reversed, the latencies decrease by \num{12}--\num{14}\,cycles (\num{6}--\unit{7}{\nano\second}) when switching from NUMA node 2 to 3 as home node.

These results highlight the importance of a careful consideration of the NUMA properties of Rome processors.
If the processor is set up to report only as a single NUMA domain to the OS, thread pinning should be used to avoid these latency penalties.

\subsection{Intel Xeon Cascade Lake SP}

We performed latency measurements on CLX at its nominal core frequency of \unit{2.5}{\giga\hertz}
 and \unit{2.4}{\giga\hertz} uncore frequency.
Latencies for cache lines in state Forward and Shared are reported as one value as they are identical on CLX.
As described by McCalpin in~\cite{McCalpinSKXTiles}, processors with a reduced core count can have different deactivated core tiles.
We used the \texttt{*\_RING\_BL\_IN\_USE} performance counters~\cite[Section 2.2.3]{IntelCASUncorePerf2017} to find the mapping of CPUs (as reported by the OS) to core tiles in our system, which we present in \figref{clx-tiles}.
While analyzing the layout, we observed that the third UPI link is not present and/or not used on our system.
This behavior was also observed by McCalpin.

\begin{figure}[b]
\centering
  \includegraphics[width=\columnwidth]{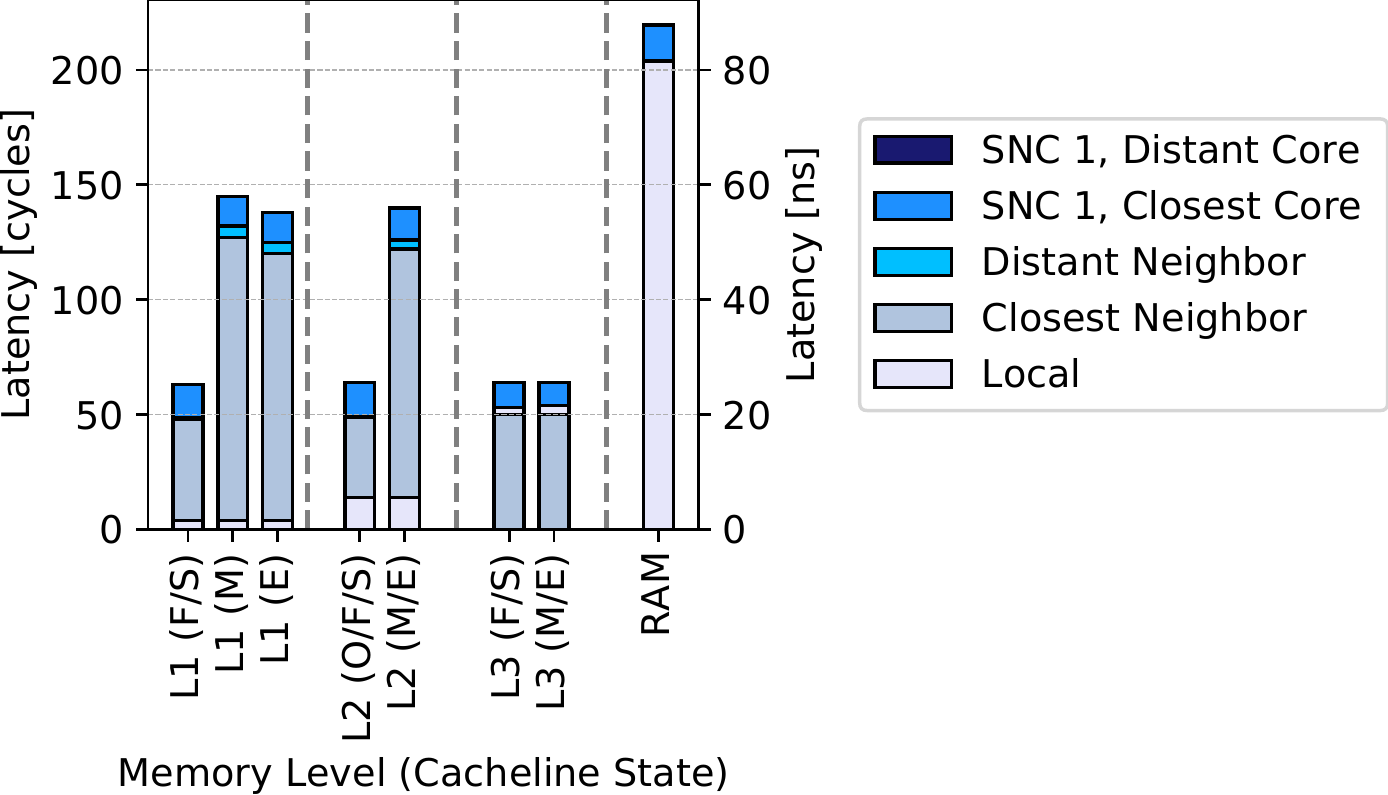}
 \caption{\label{fig:CLX-socket-latencies}Memory read latencies for an Intel Xeon Gold 6248 on one socket, measured from SNC 0 – 
     \emph{Forwarded} and \emph{Shared} states result in identical latencies, \emph{Modified} and \emph{Exclusive} latencies only for L2 and L3. Refer \tabref{CLX-latencies} for precise values.}
 \Description{Stacked bar chart of latency measured when accessing data located in remote caches depending on core locations and MOESI state of the accessed cache lines.}
\end{figure}

\begin{figure*}[t]
    \centering
    \subfigure[\label{fig:rome-2S-latencies-latencies}Socket-socket RAM-latencies for several NUMA node pairings.]{\includegraphics[height=0.18\textheight]{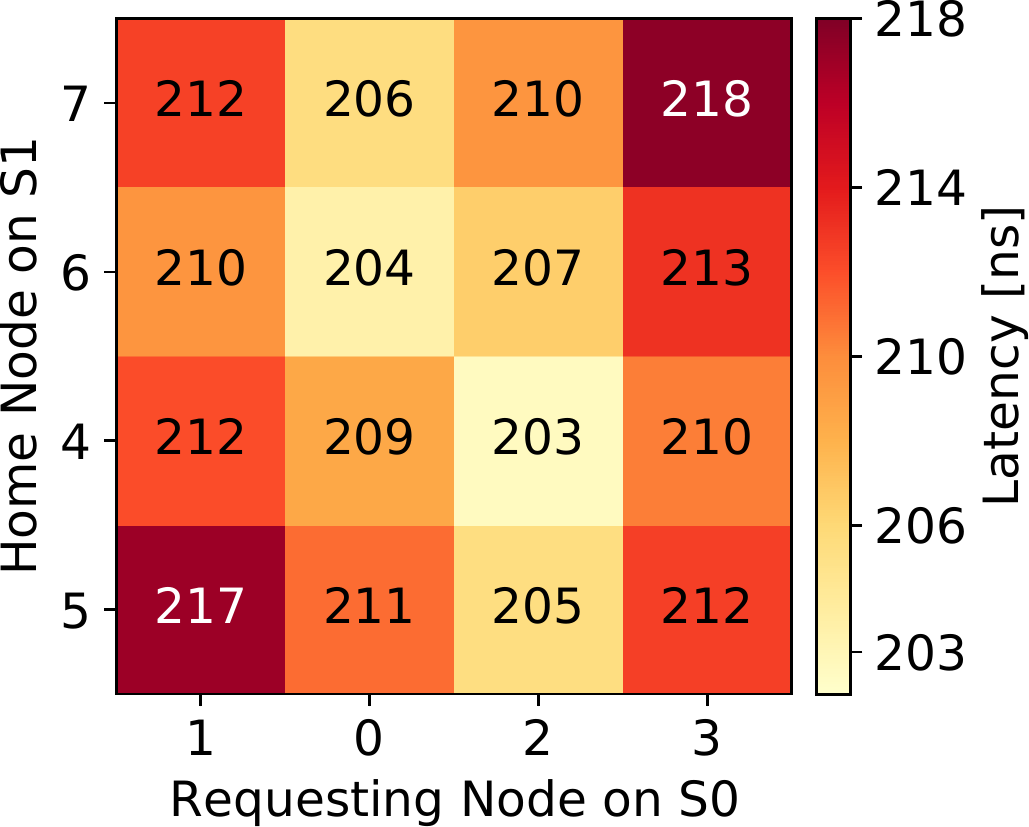}}
    \hfil
    \subfigure[\label{fig:rome-2S-latencies-io-die}Two AMD EPYC Rome processors with interconnect paths according to measurements.]{\includegraphics[height=0.18\textheight]{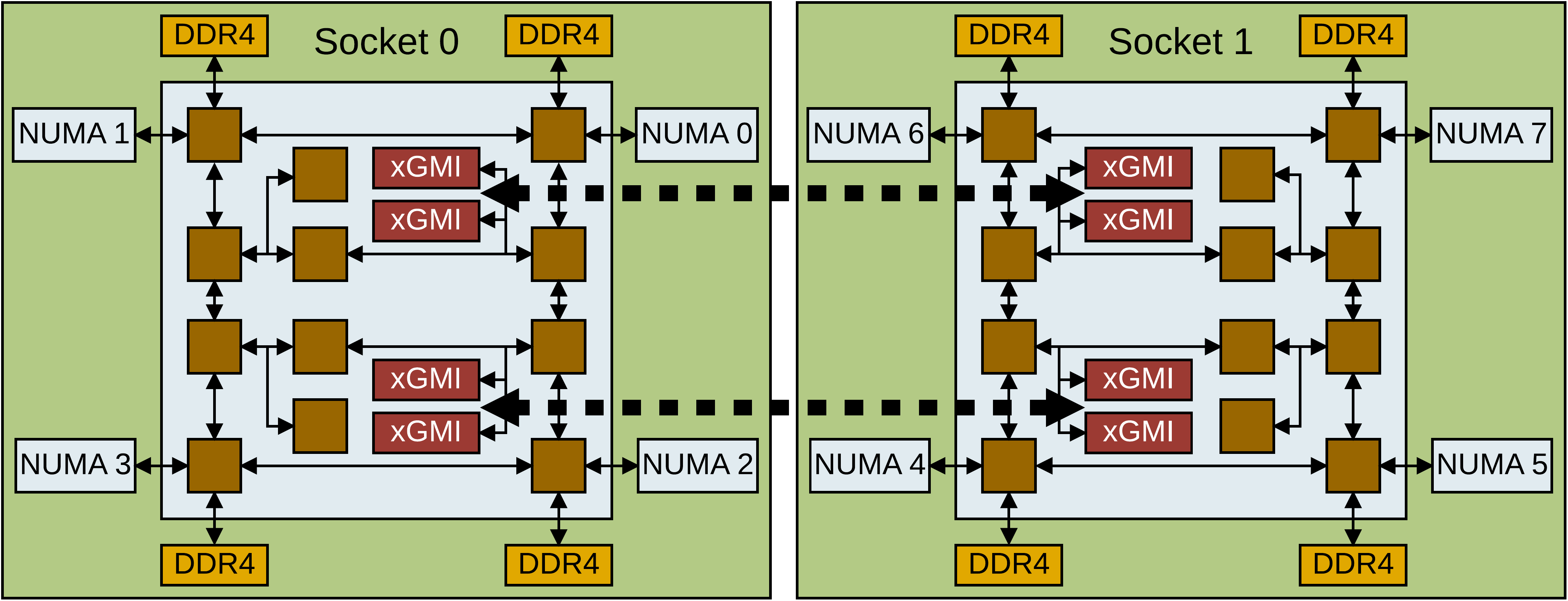}}
    \caption{\label{fig:rome-socket-socket-latencies}AMD EPYC Rome 7702: socket-socket memory latencies.
    The xGMI interfaces to a second processor on the same mainboard cannot be accessed by all cores in the same time due to their asymmetric placement in the I/O-die.
    }
    \Description{(a) Heat map for AMD EPYC Rome 7702 memory access latencies when accessing memory on a second socket.
(b) layout of how two two processors are connected via xGMI with the following simplified network (IF repeaters were removed):
socket 0 nodes: (NUMA0, NUMA1, NUMA2, NUMA3, DDR1, DDR2, DDR3, DDR4, IF-SWITCH-1, IF-SWITCH-2, IF-SWITCH-3, IF-SWITCH-4, IF-SWITCH-5, IF-SWITCH-6, IF-SWITCH-7, IF-SWITCH-8, IF-SWITCH-9, IF-SWITCH-10, IF-SWITCH-11, IF-SWITCH-12, xGMI-1, xGMI-2, xGMI-3, xGMI-4)
socket 0 edges: ( (NUMA1, IF-SWITCH-1), (DDR1, IF-SWITCH-1), (NUMA0, IF-SWITCH-2), (DDR2, IF-SWITCH-2), (IF-SWITCH-1, IF-SWITCH-2), (IF-SWITCH-1, IF-SWITCH-5), (IF-SWITCH-2, IF-SWITCH-6), (IF-SWITCH-5, IF-SWITCH-7), (IF-SWITCH-5, IF-SWITCH-8), (IF-SWITCH-7, IF-SWITCH-6), (IF-SWITCH-6, xGMI-1), (IF-SWITCH-6, xGMI-2), (NUMA3, IF-SWITCH-3), (DDR3, IF-SWITCH-3), (IF-SWITCH-3, IF-SWITCH-4), (NUMA2, IF-SWITCH-4), (DDR4, IF-SWITCH-4), (IF-SWITCH-3, IF-SWITCH-4),(IF-SWITCH-3, IF-SWITCH-9), (IF-SWITCH-4, IF-SWITCH-10), (IF-SWITCH-9, IF-SWITCH-11), (IF-SWITCH-9, IF-SWITCH-12), (IF-SWITCH-9, IF-SWITCH-10), (IF-SWITCH-10, xGMI-3), (IF-SWITCH-10, xGMI-4) )
Changes for socket 1: duplicate and rename all nodes. specific renamings: NUMA0->NUMA6, NUMA1->NUMA7, NUMA3->NUMA5, NUMA2->NUMA4, xGMI-1->xGMI-5, xGMI-2->xGMI-6, xGMI-3->xGMI-7, xGMI-4->xGMI-8.
Additional edges between sockets: ((xGMI-1, xGMI-5), (xGMI-2, xGMI-4), (xGMI-3, xGMI-7), (xGMI-4, xGMI-8))
}
\end{figure*}

\figref{CLX-socket-latencies} shows that latencies of the CLX system depend on the relative positions of the cores in the mesh, as seen in the case of latencies from core 0 to its closest and distant neighbor within the SNC.
The differences between the lowest and highest L1 and L2 latencies within an SNC can be roughly assumed to be \unit{2}{\nano\second} or \num{5}\,cycles, highlighting the effectiveness of Intel's mesh design.
Increased latencies can also be observed for accesses to memory on the second SNC. 
While L2 and L3 cache latencies for the states \emph{Modified} and \emph{Exclusive} are identical and have therefore been consolidated, L1 cache latencies from \emph{Modified} cache lines are higher than those for \emph{Exclusive} cache lines:
Reading \emph{Exclusive} cache lines from L1 seems to be identical to L2 accesses, indicating that they can be fetched from the inclusive L2 cache.
\emph{Shared} and \emph{Forwarded} cache lines are fetched from the L3 cache, if they are not present in local caches, as described in~\cite{KumarSLX}.

When accessing the second SNC on the same socket, an additional latency of \textasciitilde\num{15}\,cycles (\textasciitilde\unit{6}{\nano\second}) is incurred for all cache levels.
In contrast, accesses to other CCX on Rome add an additional \num{130}--\num{200}\,cycles.
These results indicate that CLX processors may be better suited for shared memory workloads where more than four cores work on the same data.
RAM latencies for local and remote SNC are much lower than those on Rome, as it was to be expected due to the less complex design of CLX processors.

Our results highlight the benefit of having the memory controllers integrated into the fabric mesh as it is the case with the Intel processor.
Whilst L3 latencies on CLX are \textasciitilde\num{15}\,cycles higher than on Rome, all CLX cores can access the L3 at similar latencies.
This is a potential benefit compared to Rome, where only four cores within a CCX share a L3 cache, in particular for workloads that rely on many cores that share cache lines.
If, however, a large per-core L3 is required, Rome processors may be better suited due to their larger per-core L3 slices and lower local L3 latencies.

\section{Inter-Socket Main Memory Latencies}
\label{sec:inter-socket}
While previous measurements only showed memory characteristics for accesses within a single socket, adding another processor will further increase complexity.
Again, we use BenchIT's latency benchmark to measure main memory latencies at nominal frequencies.
We schedule threads to the first core within each NUMA domain and analyze RAM latencies for all inter-socket node-pairings.

We present results for the Rome system in \figref{rome-2S-latencies-latencies}.
Here, a clear split along one axis of the processor can be observed.
The latency patterns indicate that the xGMI interfaces are not cross-connected but connected as shown in \figref{rome-2S-latencies-io-die}.
Threads scheduled on NUMA nodes that are closer to the xGMI interfaces communicate with lower latencies with the remote socket.
Communication between nodes 0 and 6 or 2 and 4, respectively, takes the least time, \num{406}--\num{408}\,cycles (\num{203}--\unit{204}{\nano\second}).
Apparently, in those cases only one IF-switch hop is required per socket in addition to the mandatory hop when accessing the I/O-die.
The highest latencies occur for ``diagonal'' accesses, e.g., from node 1 on socket 0 to node 5 on socket 1.
For those cases, \textasciitilde\num{436}\,cycles (\unit{218}{\nano\second}) are required, \num{30}\,cycles (\unit{15}{\nano\second}) more than the fastest case.

\begin{figure}[t]
\centering
  \includegraphics[width=0.6\columnwidth]{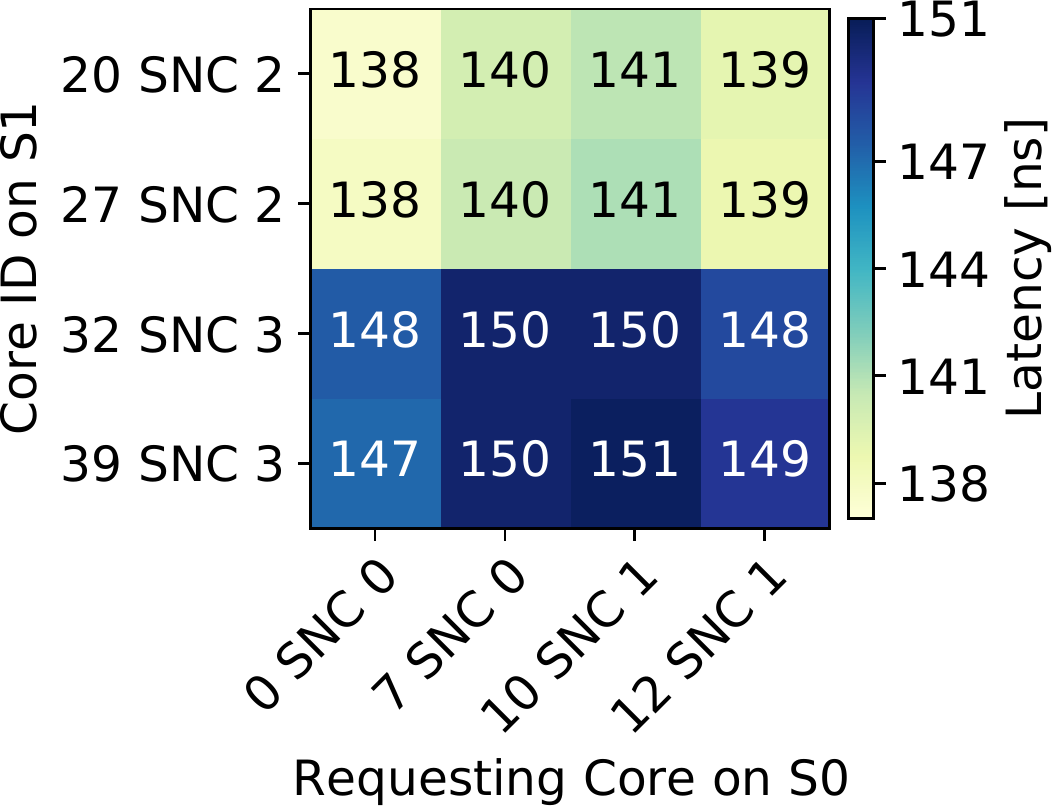}
 \caption{\label{fig:CLX-2S-latencies}Socket-Socket main memory read latencies for an Intel Xeon Gold 6248. Only the UPI-tile with two UPI links is being used.}
 \Description{Heat map for socket-socket main memory read latencies for an Intel Xeon Gold 6248. Requests oigin from different cores on SNC0 and SNC1 (socket 0), memory is located at SNC2 and SNC3 (socket 2). Accesses to SNC3 take significantly longer (~149 ns) then accesses to SNC2 (~139 ns)}
\end{figure}

Main memory accesses to the second socket on the CLX system traverse through the UPI interfaces.
In our case, only the interface tile with two UPI links is being used (see \figref{clx-tiles}), as performance counter events proved\footnote{We measured \texttt{\{R|T\}xL\_FLITS\_ALL\_DATA} events for all UPI ports for various inter-socket accesses.}.
As we show in \figref{CLX-2S-latencies}, the lowest latencies can be measured when core 0 requests memory from the first SNC on the second socket (SNC 2), independently of the core that allocated memory in this NUMA node.
At \unit{138}{\nano\second}, this value is \textasciitilde\unit{60}{\nano\second} higher than a local main memory access.
Accessing SNC 3 takes an additional \unit{10}{\nano\second}.
Based on the location of the accessing core on the first socket, additional latencies can be observed.

Our measurements show that transferring data to/from a remote socket comes at a high additional latency penalty.
If communication between to sockets is necessary, it is beneficial to schedule communicating threads carefully in order to avoid additional overhead.

\section{Summary, Conclusion, and Future Work}
\label{sec:conclusion}
In this paper, we provided in-depth descriptions of the architectures of current AMD Rome and Intel CLX server processors.
The modular chiplet design of the Rome processor allows AMD to assemble up to 64 cores in one package and ensures that all SKUs can profit from the same number of I/O interfaces and memory controllers.
This design results in accentuated NUMA properties and an L3 cache that is only shared among four cores at a time.
In contrast, Intel uses a monolithic design without inherent NUMA properties.
It scales up to 28 cores and shares a common L3 cache among all cores.

We updated the existing x86-membench code of the BenchIT suite and improved support for the cache design of these processors. 
The cache flush routines are now able to reflect the usage of inclusive L2 and exclusive L3 caches by both AMD and Intel in their current designs.
The throughput benchmark can now use AVX-512 operations with up to 32 512-bit zmm vector registers.

We also evaluated the cache and main memory latencies of the two processors.
Our measurements revealed similar latencies for local cache accesses to the L1 and L2 for both processors.
The L3 access latencies to the CCX-local L3 slices take fewer cycles on AMD processors.
However, as only four cores share the CCX-local L3, there is no commonly shared cache with low latencies for all cores on AMD Rome processors.
Therefore, shared memory workloads which rely heavily on data sharing among cores may profit from Intel's monolithic design, as all cores share a common L3 with similar latencies.

We demonstrated the complex NUMA properties of AMD's Infinity Fabric grid, both for intra-socket latencies, and inter-socket latencies.
The CLX architecture only exhibits small variances for remote cache access latencies, making these processors potentially better suited for shared memory workloads. 

More in-depth analyses of latencies for the Rome processor stress the importance of well thought out data placement when using these processors. 
This is caused by the I/O-die and the latencies it adds to data transfers that pass through it.

The L1 and L2 cache bandwidth of the AMD processor is close to its theoretical maximum for each vector length we evaluated.
This is not the case for the CLX processor.
Instead, we observed L1 bandwidths that were below the theoretical maximum for each vector length and below the sustainable bandwidths Intel defined in~\cite[Table 2-6, Section 2.2.1.3]{IntelOptimization2020}.
A similar picture emerges for the L2 cache bandwidth.
Here, we expected bandwidths similar to the L1 bandwidth for AVX and SSE instructions, as the L1 bandwidth was well within the specified L2 bandwidth.
Instead, we observed L2 bandwidths well below the results for the L1.
Our cache bandwidth results are in line with other research on these processors~\cite{Alappat2020}.

We observed a \num{41}\,\% lower RAM bandwidth for an entire socket for the CLX processors compared to Rome in our throughput benchmark.
This is a result of the lower number of memory channels of CLX processors.
Although the Intel design allows one core to access three memory channels when using the SNC configuration, a lower single core bandwidth was observed compared to AMD.
We therefore conclude that Rome processors may be better suited for memory bandwidth bound workloads than their CLX counterparts, as long as data sharing among cores is avoided as much as possible.

Future research could focus on the new architectures by Intel and AMD, Ice Lake SP and EPYC Milan.
Furthermore, the impact of our findings on real world applications should be considered in future work.
\label{sec:summary}

\begin{acks}
The authors gratefully acknowledge the compute resources and support provided by NHR@FAU for the Caskade Lake SP system. 
We used the Romeo partition by NEC of the Bull Cluster TAURUS at the Center for Information Services and High Performance Computing (ZIH) at TU Dresden for measurements on AMD Rome.
\end{acks}

\bibliographystyle{ACM-Reference-Format}

\bibliography{paper}


\begin{thebibliography}{33}


\ifx \showCODEN    \undefined \def \showCODEN     #1{\unskip}     \fi
\ifx \showDOI      \undefined \def \showDOI       #1{#1}\fi
\ifx \showISBNx    \undefined \def \showISBNx     #1{\unskip}     \fi
\ifx \showISBNxiii \undefined \def \showISBNxiii  #1{\unskip}     \fi
\ifx \showISSN     \undefined \def \showISSN      #1{\unskip}     \fi
\ifx \showLCCN     \undefined \def \showLCCN      #1{\unskip}     \fi
\ifx \shownote     \undefined \def \shownote      #1{#1}          \fi
\ifx \showarticletitle \undefined \def \showarticletitle #1{#1}   \fi
\ifx \showURL      \undefined \def \showURL       {\relax}        \fi
\providecommand\bibfield[2]{#2}
\providecommand\bibinfo[2]{#2}
\providecommand\natexlab[1]{#1}
\providecommand\showeprint[2][]{arXiv:#2}

\bibitem[\protect\citeauthoryear{{Advanced Micro Devices, Inc.}}{{Advanced
  Micro Devices, Inc.}}{2020}]%
        {AMDRomeRR2020}
\bibfield{author}{\bibinfo{person}{{Advanced Micro Devices, Inc.}}}
  \bibinfo{year}{2020}\natexlab{}.
\newblock \bibinfo{booktitle}{\emph{{Preliminary Processor Programming
  Reference (PPR) for AMD Family 17hModel 31h, Revision B0Processors}}}.
\newblock
\urldef\tempurl%
\url{https://developer.amd.com/wp-content/resources/55803_B0_PUB_0_91.pdf}
\showURL{%
\tempurl}


\bibitem[\protect\citeauthoryear{{{Advanced Micro Devices, Inc.}}}{{{Advanced
  Micro Devices, Inc.}}}{2020}]%
        {AMDOptimization2020}
\bibfield{author}{\bibinfo{person}{{{Advanced Micro Devices, Inc.}}}}
  \bibinfo{year}{2020}\natexlab{}.
\newblock \bibinfo{booktitle}{\emph{{Software Optimization Guide for AMD Family
  17h Models 30h and Greater Processors}}}.
\newblock
\urldef\tempurl%
\url{https://developer.amd.com/wp-content/resources/56305.zip}
\showURL{%
\tempurl}


\bibitem[\protect\citeauthoryear{{Akhilesh Kumar}, {Don Soltis}, {Irma Esmer},
  {Adi Yoaz}, and {Sailesh Kottapalli}}{{Akhilesh Kumar} et~al\mbox{.}}{2017}]%
        {KumarSLX}
\bibfield{author}{\bibinfo{person}{{Akhilesh Kumar}}, \bibinfo{person}{{Don
  Soltis}}, \bibinfo{person}{{Irma Esmer}}, \bibinfo{person}{{Adi Yoaz}}, {and}
  \bibinfo{person}{{Sailesh Kottapalli}}.} \bibinfo{year}{2017}\natexlab{}.
\newblock \bibinfo{title}{{The New Intel® Xeon® Processor Scalable Family
  (Formerly Skylake-SP)}}.
\newblock
\newblock
\urldef\tempurl%
\url{https://old.hotchips.org/wp-content/uploads/hc_archives/hc29/HC29.22-Tuesday-Pub/HC29.22.90-Server-Pub/HC29.22.930-Xeon-Skylake-sp-Kumar-Intel.pdf}
\showURL{%
\tempurl}
\newblock
\shownote{Hot Chips 29.}


\bibitem[\protect\citeauthoryear{{Arafa}, {Fahim}, {Kottapalli}, {Kumar},
  {Looi}, {Mandava}, {Rudoff}, {Steiner}, {Valentine}, {Vedaraman}, and
  {Vora}}{{Arafa} et~al\mbox{.}}{2019}]%
        {ArafaCLX2019}
\bibfield{author}{\bibinfo{person}{M. {Arafa}}, \bibinfo{person}{B. {Fahim}},
  \bibinfo{person}{S. {Kottapalli}}, \bibinfo{person}{A. {Kumar}},
  \bibinfo{person}{L.~P. {Looi}}, \bibinfo{person}{S. {Mandava}},
  \bibinfo{person}{A. {Rudoff}}, \bibinfo{person}{I.~M. {Steiner}},
  \bibinfo{person}{B. {Valentine}}, \bibinfo{person}{G. {Vedaraman}}, {and}
  \bibinfo{person}{S. {Vora}}.} \bibinfo{year}{2019}\natexlab{}.
\newblock \showarticletitle{{Cascade Lake: Next Generation Intel Xeon Scalable
  Processor}}.
\newblock \bibinfo{journal}{\emph{IEEE Micro}} (\bibinfo{year}{2019}).
\newblock
\urldef\tempurl%
\url{https://doi.org/10.1109/MM.2019.2899330}
\showDOI{\tempurl}


\bibitem[\protect\citeauthoryear{{Burd}, {Beck}, {White}, {Paraschou},
  {Kalyanasundharam}, {Donley}, {Smith}, {Hewitt}, and {Naffziger}}{{Burd}
  et~al\mbox{.}}{2019}]%
        {BurdZeppelin}
\bibfield{author}{\bibinfo{person}{T. {Burd}}, \bibinfo{person}{N. {Beck}},
  \bibinfo{person}{S. {White}}, \bibinfo{person}{M. {Paraschou}},
  \bibinfo{person}{N. {Kalyanasundharam}}, \bibinfo{person}{G. {Donley}},
  \bibinfo{person}{A. {Smith}}, \bibinfo{person}{L. {Hewitt}}, {and}
  \bibinfo{person}{S. {Naffziger}}.} \bibinfo{year}{2019}\natexlab{}.
\newblock \showarticletitle{{“Zeppelin”: An SoC for Multichip
  Architectures}}.
\newblock \bibinfo{journal}{\emph{IEEE Journal of Solid-State Circuits}}
  (\bibinfo{year}{2019}).
\newblock
\urldef\tempurl%
\url{https://doi.org/10.1109/JSSC.2018.2873584}
\showDOI{\tempurl}


\bibitem[\protect\citeauthoryear{{C. L. Alappat}, {J. Hofmann}, {G. Hager}, {H.
  Fehske}, {A. R. Bishop}, and {G. Wellein}}{{C. L. Alappat}
  et~al\mbox{.}}{2020}]%
        {Alappat2020}
\bibfield{author}{\bibinfo{person}{{C. L. Alappat}}, \bibinfo{person}{{J.
  Hofmann}}, \bibinfo{person}{{G. Hager}}, \bibinfo{person}{{H. Fehske}},
  \bibinfo{person}{{A. R. Bishop}}, {and} \bibinfo{person}{{G. Wellein}}.}
  \bibinfo{year}{2020}\natexlab{}.
\newblock \showarticletitle{{Understanding HPC Benchmark Performance on Intel
  Broadwell and Cascade Lake Processors}}. In
  \bibinfo{booktitle}{\emph{International Conference on High Performance
  Computing}}.
\newblock
\urldef\tempurl%
\url{https://doi.org/10.1007/978-3-030-50743-5_21}
\showDOI{\tempurl}


\bibitem[\protect\citeauthoryear{{D. Suggs}, {D. Bouvier}, {M. Clark}, {K.
  Lepak}, and {M. Subramony}}{{D. Suggs} et~al\mbox{.}}{2019}]%
        {Suggs2019}
\bibfield{author}{\bibinfo{person}{{D. Suggs}}, \bibinfo{person}{{D. Bouvier}},
  \bibinfo{person}{{M. Clark}}, \bibinfo{person}{{K. Lepak}}, {and}
  \bibinfo{person}{{M. Subramony}}.} \bibinfo{year}{2019}\natexlab{}.
\newblock \showarticletitle{{AMD “ZEN 2”}}. In
  \bibinfo{booktitle}{\emph{IEEE Hot Chips 31 Symposium (HCS)}}.
\newblock
\urldef\tempurl%
\url{https://doi.org/10.1109/HOTCHIPS.2019.8875673}
\showDOI{\tempurl}


\bibitem[\protect\citeauthoryear{{D. Suggs}, {M. Subramony}, and {D.
  Bouvier}}{{D. Suggs} et~al\mbox{.}}{2020}]%
        {Suggs2020}
\bibfield{author}{\bibinfo{person}{{D. Suggs}}, \bibinfo{person}{{M.
  Subramony}}, {and} \bibinfo{person}{{D. Bouvier}}.}
  \bibinfo{year}{2020}\natexlab{}.
\newblock \showarticletitle{{The AMD “Zen 2” Processor}}.
\newblock \bibinfo{journal}{\emph{IEEE Micro}} (\bibinfo{year}{2020}).
\newblock
\urldef\tempurl%
\url{https://doi.org/10.1109/MM.2020.2974217}
\showDOI{\tempurl}


\bibitem[\protect\citeauthoryear{Fog}{Fog}{2020}]%
        {AgnerMicroarchitecture}
\bibfield{author}{\bibinfo{person}{A. Fog}.} \bibinfo{year}{2020}\natexlab{}.
\newblock \bibinfo{title}{{3. The microarchitecture of Intel, AMD and VIA CPUs:
  An optimization guide for assembly programmers and compiler makers}}.
\newblock
\newblock
\urldef\tempurl%
\url{https://www.agner.org/optimize/microarchitecture.pdf}
\showURL{%
\tempurl}
\newblock
\shownote{Technical University of Denmark.}


\bibitem[\protect\citeauthoryear{Hackenberg, Sch{\"{o}}ne, Ilsche, Molka,
  Schuchart, and Geyer}{Hackenberg et~al\mbox{.}}{2015}]%
        {Hackenberg2015_Haswell}
\bibfield{author}{\bibinfo{person}{Daniel Hackenberg}, \bibinfo{person}{Robert
  Sch{\"{o}}ne}, \bibinfo{person}{Thomas Ilsche}, \bibinfo{person}{Daniel
  Molka}, \bibinfo{person}{Joseph Schuchart}, {and} \bibinfo{person}{Robin
  Geyer}.} \bibinfo{year}{2015}\natexlab{}.
\newblock \showarticletitle{{An Energy Efficiency Feature Survey of the {Intel}
  {Haswell} Processor}}. In \bibinfo{booktitle}{\emph{International Parallel
  and Distributed Processing Symposium Workshop}}.
\newblock
\urldef\tempurl%
\url{https://doi.org/10.1109/ipdpsw.2015.70}
\showDOI{\tempurl}


\bibitem[\protect\citeauthoryear{{Intel Corporation}}{{Intel
  Corporation}}{[n.d.]}]%
        {IntelCLXBrief}
\bibfield{author}{\bibinfo{person}{{Intel Corporation}}.}
  \bibinfo{year}{[n.d.]}\natexlab{}.
\newblock \bibinfo{title}{{Product brief - Intel Xeon Scalable Platform –
  Second Generation - Intel Xeon Scalable Processors}}.
\newblock
\newblock
\urldef\tempurl%
\url{https://www.intel.com/content/dam/www/public/us/en/documents/product-briefs/2nd-gen-xeon-scalable-processors-brief-Feb-2020-2.pdf}
\showURL{%
\tempurl}


\bibitem[\protect\citeauthoryear{{{Intel Corporation}}}{{{Intel
  Corporation}}}{2017}]%
        {IntelCASUncorePerf2017}
\bibfield{author}{\bibinfo{person}{{{Intel Corporation}}}.}
  \bibinfo{year}{2017}\natexlab{}.
\newblock \bibinfo{booktitle}{\emph{{Intel Xeon Processor Scalable Memory
  Family Uncore Performance Monitoring}}}.
\newblock
\urldef\tempurl%
\url{https://www.intel.com/content/www/us/en/processors/xeon/scalable/xeon-scalable-uncore-performance-monitoring-manual.html}
\showURL{%
\tempurl}


\bibitem[\protect\citeauthoryear{{{Intel Corporation}}}{{{Intel
  Corporation}}}{2020a}]%
        {IntelSkylakeSpec}
\bibfield{author}{\bibinfo{person}{{{Intel Corporation}}}.}
  \bibinfo{year}{2020}\natexlab{a}.
\newblock \bibinfo{booktitle}{\emph{{Intel® Xeon® Processor Scalable Family
  – Specification Update}}}.
\newblock
\urldef\tempurl%
\url{https://www.intel.com/content/dam/www/public/us/en/documents/specification-updates/xeon-scalable-spec-update.pdf}
\showURL{%
\tempurl}


\bibitem[\protect\citeauthoryear{{{Intel Corporation}}}{{{Intel
  Corporation}}}{2020b}]%
        {IntelCLXSpecUpdate}
\bibfield{author}{\bibinfo{person}{{{Intel Corporation}}}.}
  \bibinfo{year}{2020}\natexlab{b}.
\newblock \bibinfo{booktitle}{\emph{{Second Generation Intel® Xeon® Scalable
  Processors – Specification Update}}}.
\newblock
\urldef\tempurl%
\url{https://www.intel.com/content/www/au/en/products/docs/processors/xeon/2nd-gen-xeon-scalable-spec-update.html}
\showURL{%
\tempurl}


\bibitem[\protect\citeauthoryear{{{Intel Corporation}}}{{{Intel
  Corporation}}}{2021}]%
        {IntelOptimization2020}
\bibfield{author}{\bibinfo{person}{{{Intel Corporation}}}.}
  \bibinfo{year}{2021}\natexlab{}.
\newblock \bibinfo{booktitle}{\emph{{Intel® 64 and IA-32 Architectures
  Optimization Reference Manual}}}.
\newblock
\urldef\tempurl%
\url{https://software.intel.com/content/dam/develop/external/us/en/documents-tps/64-ia-32-architectures-optimization-manual.pdf}
\showURL{%
\tempurl}


\bibitem[\protect\citeauthoryear{{John McCalpin}}{{John McCalpin}}{2018}]%
        {McCalpinCacheHash}
\bibfield{author}{\bibinfo{person}{{John McCalpin}}.}
  \bibinfo{year}{2018}\natexlab{}.
\newblock \bibinfo{title}{{Address Hashing in Intel Processors)}}.
\newblock
\newblock
\urldef\tempurl%
\url{https://doi.org/10.26153/tsw/13161}
\showDOI{\tempurl}


\bibitem[\protect\citeauthoryear{{John McCalpin}}{{John McCalpin}}{2021}]%
        {McCalpinSKXTiles}
\bibfield{author}{\bibinfo{person}{{John McCalpin}}.}
  \bibinfo{year}{2021}\natexlab{}.
\newblock \bibinfo{title}{{Mapping Core and L3 Slice Numbering to Die Location
  in Intel Xeon Scalable Processors)}}.
\newblock
\newblock
\urldef\tempurl%
\url{https://doi.org/10.26153/tsw/13119}
\showDOI{\tempurl}


\bibitem[\protect\citeauthoryear{Juckeland, Börner, Kluge, Kölling, Nagel,
  Pflüger, Röding, Seidl, William, and Wloch}{Juckeland
  et~al\mbox{.}}{2004}]%
        {JuckelandBenchIT}
\bibfield{author}{\bibinfo{person}{G. Juckeland}, \bibinfo{person}{S. Börner},
  \bibinfo{person}{M. Kluge}, \bibinfo{person}{S. Kölling},
  \bibinfo{person}{W.E. Nagel}, \bibinfo{person}{S. Pflüger},
  \bibinfo{person}{H. Röding}, \bibinfo{person}{S. Seidl}, \bibinfo{person}{T.
  William}, {and} \bibinfo{person}{R. Wloch}.} \bibinfo{year}{2004}\natexlab{}.
\newblock \showarticletitle{{BenchIT — Performance measurement and comparison
  for scientific applications}}.
\newblock In \bibinfo{booktitle}{\emph{Parallel Computing}}.
\newblock
\urldef\tempurl%
\url{https://doi.org/10.1016/S0927-5452(04)80064-9}
\showDOI{\tempurl}


\bibitem[\protect\citeauthoryear{Kashyap}{Kashyap}{2020}]%
        {Kashyap2020}
\bibfield{author}{\bibinfo{person}{A. Kashyap}.}
  \bibinfo{year}{2020}\natexlab{}.
\newblock \bibinfo{booktitle}{\emph{{High Performance Computing: Tuning Guide
  for AMD EPYC™ 7002 Series Processors}}}.
\newblock {Advanced Micro Devices, Inc}.
\newblock
\urldef\tempurl%
\url{https://developer.amd.com/wp-content/resources/56827-1-0.pdf}
\showURL{%
\tempurl}


\bibitem[\protect\citeauthoryear{Molka}{Molka}{2017}]%
        {MolkaDiss}
\bibfield{author}{\bibinfo{person}{Daniel Molka}.}
  \bibinfo{year}{2017}\natexlab{}.
\newblock \emph{\bibinfo{title}{{Performance Analysis of Complex Shared Memory
  Systems}}}.
\newblock \bibinfo{thesistype}{Ph.D. Dissertation}. \bibinfo{school}{Technische
  Universität Dresden}, \bibinfo{address}{Dresden}.
\newblock
\urldef\tempurl%
\url{https://nbn-resolving.org/urn:nbn:de:bsz:14-qucosa-221729}
\showURL{%
\tempurl}


\bibitem[\protect\citeauthoryear{Molka, Hackenberg, and Sch\"{o}ne}{Molka
  et~al\mbox{.}}{2014}]%
        {MolkaSandybridge}
\bibfield{author}{\bibinfo{person}{Daniel Molka}, \bibinfo{person}{Daniel
  Hackenberg}, {and} \bibinfo{person}{Robert Sch\"{o}ne}.}
  \bibinfo{year}{2014}\natexlab{}.
\newblock \showarticletitle{{Main Memory and Cache Performance of Intel Sandy
  Bridge and AMD Bulldozer}}. In \bibinfo{booktitle}{\emph{Workshop on Memory
  Systems Performance and Correctness (MSPC)}}.
\newblock
\urldef\tempurl%
\url{https://doi.org/10.1145/2618128.2618129}
\showDOI{\tempurl}


\bibitem[\protect\citeauthoryear{{Molka}, {Hackenberg}, {Schöne}, and
  {Müller}}{{Molka} et~al\mbox{.}}{2009}]%
        {MolkaNehalem}
\bibfield{author}{\bibinfo{person}{D. {Molka}}, \bibinfo{person}{D.
  {Hackenberg}}, \bibinfo{person}{R. {Schöne}}, {and} \bibinfo{person}{M.~S.
  {Müller}}.} \bibinfo{year}{2009}\natexlab{}.
\newblock \showarticletitle{{Memory Performance and Cache Coherency Effects on
  an Intel Nehalem Multiprocessor System}}. In \bibinfo{booktitle}{\emph{18th
  International Conference on Parallel Architectures and Compilation Techniques
  (PACT)}}.
\newblock
\urldef\tempurl%
\url{https://doi.org/10.1109/PACT.2009.22}
\showDOI{\tempurl}


\bibitem[\protect\citeauthoryear{Molka, Hackenberg, Schöne, and Nagel}{Molka
  et~al\mbox{.}}{2015}]%
        {MolkaHaswell}
\bibfield{author}{\bibinfo{person}{Daniel Molka}, \bibinfo{person}{Daniel
  Hackenberg}, \bibinfo{person}{Robert Schöne}, {and}
  \bibinfo{person}{Wolfgang~E. Nagel}.} \bibinfo{year}{2015}\natexlab{}.
\newblock \showarticletitle{{Cache Coherence Protocol and Memory Performance of
  the Intel Haswell-EP Architecture}}. In \bibinfo{booktitle}{\emph{44th
  International Conference on Parallel Processing (ICPP)}}.
\newblock
\urldef\tempurl%
\url{https://doi.org/10.1109/ICPP.2015.83}
\showDOI{\tempurl}


\bibitem[\protect\citeauthoryear{Mulnix}{Mulnix}{2017}]%
        {MulnixSLX2017}
\bibfield{author}{\bibinfo{person}{D. Mulnix}.}
  \bibinfo{year}{2017}\natexlab{}.
\newblock \bibinfo{title}{{Intel® Xeon® Processor Scalable Family Technical
  Overview}}.
\newblock
\newblock
\urldef\tempurl%
\url{https://software.intel.com/en-us/articles/intel-xeon-processor-scalable-family-technical-overview}
\showURL{%
\tempurl}


\bibitem[\protect\citeauthoryear{Naffziger, Beck, Burd, Lepak, Loh, Subramony,
  and White}{Naffziger et~al\mbox{.}}{2021}]%
        {Naffziger2021}
\bibfield{author}{\bibinfo{person}{Samuel Naffziger}, \bibinfo{person}{Noah
  Beck}, \bibinfo{person}{Thomas Burd}, \bibinfo{person}{Kevin Lepak},
  \bibinfo{person}{Gabriel~H. Loh}, \bibinfo{person}{Mahesh Subramony}, {and}
  \bibinfo{person}{Sean White}.} \bibinfo{year}{2021}\natexlab{}.
\newblock \showarticletitle{Pioneering Chiplet Technology and Design for the
  AMD EPYC™ and Ryzen™ Processor Families : Industrial Product}. In
  \bibinfo{booktitle}{\emph{2021 ACM/IEEE 48th Annual International Symposium
  on Computer Architecture (ISCA)}}. \bibinfo{pages}{57--70}.
\newblock
\urldef\tempurl%
\url{https://doi.org/10.1109/ISCA52012.2021.00014}
\showDOI{\tempurl}


\bibitem[\protect\citeauthoryear{{Naffziger}, {Lepak}, {Paraschou}, and
  {Subramony}}{{Naffziger} et~al\mbox{.}}{2020}]%
        {Naffziger2020Paper}
\bibfield{author}{\bibinfo{person}{S. {Naffziger}}, \bibinfo{person}{K.
  {Lepak}}, \bibinfo{person}{M. {Paraschou}}, {and} \bibinfo{person}{M.
  {Subramony}}.} \bibinfo{year}{2020}\natexlab{}.
\newblock \showarticletitle{{2.2 AMD Chiplet Architecture for High-Performance
  Server and Desktop Products}}. In \bibinfo{booktitle}{\emph{International
  Solid- State Circuits Conference (ISSCC)}}.
\newblock
\urldef\tempurl%
\url{https://doi.org/10.1109/ISSCC19947.2020.9063103}
\showDOI{\tempurl}


\bibitem[\protect\citeauthoryear{Ramos and Hoefler}{Ramos and Hoefler}{2013}]%
        {Sabela2013}
\bibfield{author}{\bibinfo{person}{Sabela Ramos} {and} \bibinfo{person}{Torsten
  Hoefler}.} \bibinfo{year}{2013}\natexlab{}.
\newblock \showarticletitle{{Modeling Communication in Cache-Coherent SMP
  Systems: A Case-Study with Xeon Phi}}. In \bibinfo{booktitle}{\emph{22nd
  International Symposium on High-Performance Parallel and Distributed
  Computing (HPDC)}}.
\newblock
\urldef\tempurl%
\url{https://doi.org/10.1145/2462902.2462916}
\showDOI{\tempurl}


\bibitem[\protect\citeauthoryear{Sch{\"{o}}ne, Ilsche, Bielert, Gocht, and
  Hackenberg}{Sch{\"{o}}ne et~al\mbox{.}}{2019}]%
        {Schoene2019}
\bibfield{author}{\bibinfo{person}{Robert Sch{\"{o}}ne},
  \bibinfo{person}{Thomas Ilsche}, \bibinfo{person}{Mario Bielert},
  \bibinfo{person}{Andreas Gocht}, {and} \bibinfo{person}{Daniel Hackenberg}.}
  \bibinfo{year}{2019}\natexlab{}.
\newblock \showarticletitle{{Energy Efficiency Features of the Intel Skylake-SP
  Processor and Their Impact on Performance}}. In
  \bibinfo{booktitle}{\emph{International Conference on High Performance
  Computing \& Simulation (HPCS)}}.
\newblock
\urldef\tempurl%
\url{https://doi.org/10.1109/HPCS48598.2019.9188239}
\showDOI{\tempurl}


\bibitem[\protect\citeauthoryear{Sch\"{o}ne, Ilsche, Bielert, Velten, Schmidl,
  and Hackenberg}{Sch\"{o}ne et~al\mbox{.}}{2021}]%
        {romeee}
\bibfield{author}{\bibinfo{person}{Robert Sch\"{o}ne}, \bibinfo{person}{Thomas
  Ilsche}, \bibinfo{person}{Mario Bielert}, \bibinfo{person}{Markus Velten},
  \bibinfo{person}{Markus Schmidl}, {and} \bibinfo{person}{Daniel Hackenberg}.}
  \bibinfo{year}{2021}\natexlab{}.
\newblock \showarticletitle{Energy Efficiency Aspects of the AMD Zen 2
  Architecture}. In \bibinfo{booktitle}{\emph{2021 IEEE International
  Conference on Cluster Computing (CLUSTER)}}. \bibinfo{pages}{562--571}.
\newblock
\urldef\tempurl%
\url{https://doi.org/10.1109/Cluster48925.2021.00087}
\showDOI{\tempurl}


\bibitem[\protect\citeauthoryear{Strohmaier, Dongarra, Simon, Meuer, and
  Meuer}{Strohmaier et~al\mbox{.}}{2021}]%
        {Top500}
\bibfield{author}{\bibinfo{person}{Erich Strohmaier}, \bibinfo{person}{Jack
  Dongarra}, \bibinfo{person}{Horst Simon}, \bibinfo{person}{Martin Meuer},
  {and} \bibinfo{person}{Hans Meuer}.} \bibinfo{year}{2021}\natexlab{}.
\newblock \bibinfo{title}{TOP500}.
\newblock
\newblock
\newblock
\shownote{\url{https://top500.org} (accessed 2022-01-07).}


\bibitem[\protect\citeauthoryear{{Tam}, {Muljono}, {Huang}, {Iyer}, {Royneogi},
  {Satti}, {Qureshi}, {Chen}, {Wang}, {Hsieh}, {Vora}, and {Wang}}{{Tam}
  et~al\mbox{.}}{2018}]%
        {Tam2018}
\bibfield{author}{\bibinfo{person}{S.~M. {Tam}}, \bibinfo{person}{H.
  {Muljono}}, \bibinfo{person}{M. {Huang}}, \bibinfo{person}{S. {Iyer}},
  \bibinfo{person}{K. {Royneogi}}, \bibinfo{person}{N. {Satti}},
  \bibinfo{person}{R. {Qureshi}}, \bibinfo{person}{W. {Chen}},
  \bibinfo{person}{T. {Wang}}, \bibinfo{person}{H. {Hsieh}},
  \bibinfo{person}{S. {Vora}}, {and} \bibinfo{person}{E. {Wang}}.}
  \bibinfo{year}{2018}\natexlab{}.
\newblock \showarticletitle{{SkyLake-SP: A 14nm 28-Core Xeon® processor}}. In
  \bibinfo{booktitle}{\emph{International Solid - State Circuits Conference
  (ISSCC)}}.
\newblock
\urldef\tempurl%
\url{https://doi.org/10.1109/ISSCC.2018.8310170}
\showDOI{\tempurl}


\bibitem[\protect\citeauthoryear{Treibig, Hager, and Wellein}{Treibig
  et~al\mbox{.}}{2010}]%
        {likwid}
\bibfield{author}{\bibinfo{person}{J. Treibig}, \bibinfo{person}{G. Hager},
  {and} \bibinfo{person}{G. Wellein}.} \bibinfo{year}{2010}\natexlab{}.
\newblock \showarticletitle{{LIKWID: A lightweight performance-oriented tool
  suite for x86 multicore environments}}. In \bibinfo{booktitle}{\emph{First
  International Workshop on Parallel Software Tools and Tool Infrastructures
  (PSTI)}}.
\newblock
\urldef\tempurl%
\url{https://doi.org/10.1109/ICPPW.2010.38}
\showDOI{\tempurl}


\bibitem[\protect\citeauthoryear{Yarom and Falkner}{Yarom and Falkner}{2014}]%
        {Yarom2014}
\bibfield{author}{\bibinfo{person}{Yuval Yarom} {and} \bibinfo{person}{Katrina
  Falkner}.} \bibinfo{year}{2014}\natexlab{}.
\newblock \showarticletitle{{FLUSH+ RELOAD: A high resolution, low noise, L3
  cache side-channel attack}}. In \bibinfo{booktitle}{\emph{23rd {USENIX}
  Security Symposium}}.
\newblock
\showISBNx{978-1-931971-15-7}
\urldef\tempurl%
\url{https://www.usenix.org/system/files/conference/usenixsecurity14/sec14-paper-yarom.pdf}
\showURL{%
\tempurl}


\end{thebibliography}

\balance
\appendix
\section{Cache and Main Memory Latencies - Summary}\label{app:lat-summary}
\begin{table}[h!]
\centering
\caption{AMD EPYC 7702 – memory read latencies: memory accesses on one socket. \emph{Shared} cachelines fetched from another CCX are read from RAM, therefore they are listed for local CCX reads only. \emph{Invalid} cache lines are fetched from RAM at all times.}
{\setlength{\tabcolsep}{0.23em}
\begin{tabular}{lccccc}
 \hline
 & & \multicolumn{4}{c}{\textbf{Latency in [ns] ([cycles])}}\\
 \textbf{Source} & \textbf{State} & \textbf{L1} & \textbf{L2} & \textbf{L3} & \textbf{RAM} \\
\hline
Local & M/O/E/S & 2 (4) & 6 (12) & \multirow{3}{*}{19.5 (39)} & \multirow{5}{*}{110 (220)} \\
\multirow{2}{1cm}{Same CCX} & M/E & 39 (78) & 37 (74) & &\\
& O/S & 36 (72)& 37 (74)&\\
\multirow{2}{1cm}{Same CCD} & M/E & 123.5 (247) & 126.5 (253) & 120.5 (241)&\\
& O/S & 102.5 (205) & 108 (216) & 108 (216) &\\
\hline
\multirow{2}{*}{NUMA 1} & M/E & 129 (258) & 131.5 (263) & 125.5 (251) & \multirow{2}{*}{ 115 (230)}\\
& O/S & 107 (214) & 114 (228) & 115 (230) &\\
\hline
\multirow{2}{*}{NUMA 2} & M/E & 142 (284) & 145.5 (291) & 140 (280) & \multirow{2}{*}{144 (248)}\\
& O/S & 115.5 (231) & 122.5 (245) & 143 (246) &\\
\hline
\multirow{2}{*}{NUMA 3} & M/E & 148.5 (297) & 152 (304) & 146 (292) & \multirow{2}{*}{127.5 (255)}\\
& O/S & 120 (240) & 126 (252) & 151.5 (253) &\\
\hline
\end{tabular}
}
\label{tab:rome-1S-latencies}
\end{table}

\begin{table}[h!]
    \centering
    \caption{Intel Xeon Gold 6248 – memory read latencies: memory accesses on one socket. \emph{Modified} and \emph{Exclusive} cache lines have the same latencies, except for remote L1 accesses. \emph{Shared} and \emph{Forwarded} cache lines can be read with identical latencies. \emph{Invalid} cache lines are fetched from RAM at all times.}
    {\setlength{\tabcolsep}{0.1em}
    \begin{tabular}{lccccc}
    \hline
    & & \multicolumn{4}{c}{\textbf{Latency in [ns] ([cycles])}}\\
    \textbf{Source} & \textbf{State} & \textbf{L1} & \textbf{L2} & \textbf{L3} & \textbf{RAM}\\
    \hline
    Local & M/E/S/F & 1.6 (4) & 5.6 (14) & \multirow{4}{*}{21.6 (54)} & \multirow{4}{*}{80 (200)} \\
    \multirow{3}{*}{\shortstack{Same\\ SNC}} & \shortstack{M\\ \emph{ }} & \shortstack{50.8--52.8\\ (127--132)} & \multirow{2}{2cm}{\centering\shortstack{48.8--50.4\\ (122--126)}} & &\\
    & \shortstack{E\\ \emph{ }} & \shortstack{48--50\\ (120--125)} &&\\
    & S/F & \multicolumn{2}{c}{20 (50)} & & \\
    \hline
    \multirow{3}{*}{\shortstack{Other\\ SNC}} & \shortstack{M\\ \emph{ }} & \shortstack{56.4--58\\ (141--145)} & \shortstack{54.4--56\\ (136--140)}&\multirow{3}{*}{25.6 (64)} &  \multirow{3}{*}{86.4 (216)} \\
    & \shortstack{E\\ \emph{ }} & \shortstack{53.6--55.2\\ (134--138)} &\shortstack{54--56\\ (135--140)}&\\
    & S/F & \multicolumn{2}{c}{25.6 (64)}&& \\
    \hline
    \end{tabular}
    }
    \label{tab:CLX-latencies}
\end{table}

\end{document}